\documentclass{aa}  

\usepackage{xcolor}
\usepackage{hyperref}
\usepackage{threeparttable}
\usepackage{caption}
\usepackage{subcaption}
\usepackage{graphicx}
\usepackage{txfonts}
\usepackage{enumitem}

\def\mdot{$\dot{M}\rm_{ufo}$}
\def\ekin{$\dot{E}\rm_{K,ufo}$}
\def\vufo{$\rm v_{ufo}$}
\def\rufo{$R\rm_{ufo}$}
\def\lbol{$L\rm_{Bol}$}
\def\NH{$N\rm_{H}$}
\def\Mbh{$M\rm_{BH}$}
\def\ledd{$L\rm_{Edd}$}
\def\Edd{$\lambda\rm_{Edd}$}
\def\routA{$R\rm_{out}^{A}$}
\def\routBC{$R\rm_{out}^{B,C}$}
\def\trest{$\Delta$t$\rm_{rest}$}
\def\routminA{$R\rm_{out,min}^{A}$}
\def\routminBC{$R\rm_{out,min}^{B,C}$}
\def\ergs{erg s$\rm^{-1}$}
\def\vmax{$\rm v_{max}$}
\def\vmin{$\rm v_{min}$}
\def\cmdue{cm$\rm^{-2}$}
\def\cmtre{cm$\rm^{-3}$}
\def\suno{s$\rm^{-1}$}

\begin{document}

   \title{The WISSH quasars project}
    \subtitle{X. Discovery of a multi-component and highly-variable UV ultra-fast outflow in a z=3.6 quasar}
\date{}

\author{{G. Vietri}\thanks{Corresponding author: giustina.vietri@inaf.it}\inst{\ref{inst1},\ref{inst3}}
   \and
{T. Misawa}\inst{\ref{inst2}}
\and
{E. Piconcelli}\inst{\ref{inst1}}
\and
{P. Franzetti}\inst{\ref{inst3}}
\and
{A. Luminari}\inst{\ref{inst6},\ref{inst1}}
\and
{A. Travascio}\inst{\ref{inst4}}
\and
{M. Bischetti}\inst{\ref{inst5}}
\and
{S. Bisogni}\inst{\ref{inst3}}
\and
{A. Bongiorno}\inst{\ref{inst1}}
\and
{G. Bruni}\inst{\ref{inst6}}
\and
{C. Feruglio}\inst{\ref{inst5}}
\and
{A. Giunta}\inst{\ref{inst7}}
\and
{F. Nicastro}\inst{\ref{inst1}}
\and
{I. Saccheo}\inst{\ref{inst1}}
\and
{V. Testa}\inst{\ref{inst1}}
\and
{F. Tombesi}\inst{\ref{inst8},\ref{inst1},\ref{inst10},\ref{inst11},\ref{inst12}}
\and
{C. Vignali}\inst{\ref{inst9}}
\and
{L. Zappacosta}\inst{\ref{inst1}}
\and
{F. Fiore}\inst{\ref{inst5}}
}

\institute{Osservatorio Astronomico di Roma (INAF), via Frascati 33, 00040 Monte Porzio Catone (Roma), Italy\label{inst1}
           \and
        INAF - Istituto di Astrofisica Spaziale e Fisica cosmica Milano, Via Alfonso Corti 12, 20133, Milano, Italy\label{inst3}
           \and
           School of General Education, Shinshu University, 3-1-1 Asahi, Matsumoto, Nagano 390-8621, Japan\label{inst2}
           \and
        INAF - Istituto di Astrofisica e Planetologia Spaziali, Via del Fosso del Cavaliere, 100, 00133, Rome, Italy\label{inst6}
          \and
      Dipartimento di Fisica 'G. Occhialini', Università degli Studi di Milano-Bicocca, Piazza della Scienza 3, I-20126 Milano, Italy\label{inst4}
           \and
           INAF - Osservatorio Astronomico di Trieste, Via G. B. Tiepolo 11, 34143, Trieste, Italy\label{inst5}
          \and
          SSDC-ASI, Rome, Italy; Agenzia Spaziale Italiana (ASI), Rome, Italy\label{inst7}
                    \and
        Department of Physics, University of Rome `Tor Vergata', Via della Ricerca Scientifica 1, I-00133 Rome, Italy\label{inst8}
          \and
          INFN - Roma Tor Vergata, Via della Ricerca Scientifica 1, 00133, Rome, Italy\label{inst10}
          \and
          Department of Astronomy, University of Maryland, College Park, MD 20742, USA\label{inst11}
          \and
          NASA Goddard Space Flight Center, Greenbelt, MD 20771, USA\label{inst12}
          \and
          Dipartimento di Fisica e Astronomia, Università degli Studi di Bologna, Via Gobetti 93/2, 40129, Bologna, Italy\label{inst9}}

 
  \abstract
   {We report on the variability of a multi-component broad absorption line (BAL) system observed in the hyper-luminous quasar J1538+0855 at z=3.6. Observations from SDSS, VLT, LBT and Subaru telescopes taken at five different epochs, spanning 17 yr in the observed frame, are presented. We detect three (A, B, C) CIV variable troughs exhibiting extreme velocities ($\sim$40,000-54,000 km s$^{-1}$) similar to the ultra-fast outflows (UFOs) typically observed in the X-ray spectra. The A component of the BAL UFO (\vufo$\sim$0.17\textit{c}) shows strength variations, while B (\vufo$\sim$0.15\textit{c}) and C (\vufo$\sim$0.13\textit{c}) components show  changes both in shape and strength, appearing and disappearing at different epochs. In addition, during the last observation on June 2021 the entire BAL system disappears. The variability trends observed during the first two epochs (1.30 yr rest-frame) in the CIV, SiIV, OVI and NV absorption spectral regions are the same for B and C troughs, while the A component of the BAL varies independently. This suggests a change in the ionization state of the absorbing gas for B and C components and tangential motion for the A component, as causes of this temporal behavior. Accordingly, it is possible to provide an upper limit for distance of the gas responsible for the A component of \routA$\le$58 pc, and in turn, a kinetic power of \ekin$\le$5.2 $\times$ 10$^{44}$ \ergs. We also obtain \routBC$\le$2.7 kpc for B and C components, which implies an upper limit estimation of \ekin$\le$2.1$\times$10$^{46}$ \ergs\ and \ekin$\le$1.4$\times$10$^{46}$ \ergs, respectively. 
   
Future spectral monitoring with high-resolution instruments is mandatory to accurately constrain physical properties of the BAL UFO discovered in the UV spectrum of J1538+0855 and investigate its role as a promising mechanism for the origin of the extended ($\sim$75 kpc) CIV nebula surrounding this hyper-luminous quasar.}

   \maketitle
%

\section{Introduction}

There is a general consensus in regarding outflows powered by quasars (QSOs) as one of the most promising mechanisms to regulate the evolution of the massive galaxies by depositing energy, momentum and metals into the interstellar and circum-galactic medium (\citealt{Silk1998}; \citealt{Fabian2012}; \citealt{Choi2018}). Many theoretical works suggest the existence of a two-stage mechanism (\citealt{Zubovas2012}; \citealt{Faucher2012}). Radiative forces and/or magneto-centrifugal forces accelerate out from the immediate vicinity of the accretion disk a very fast wind (\citealt{Proga2007}; \citealt{Fukumura2015}) that then shocks against the ISM and accelerates the swept-up gas, thus producing the galactic-scale, massive outflows observed in the neutral/molecular gas component (e.g. \citealt{Tombesi2015},  \citealt{Bischetti2019},\citealt{Smith2019}, \citealt{Veilleux2020}). However, the study of this phenomenon is very complex as it involves multi-phase gas over scales ranging from few gravitational radii up to tens of kpc (e.g. \citealt{Cicone2018}, \citealt{Arav2018}, \citealt{Harrison2018}), and our understanding is still far from being complete.

Outflows originating from the inner regions around SMBHs are detected in a substantial fraction of AGNs ($\sim$50\%) as blueshifted absorption lines in the UV and X-ray spectra of luminous AGN (\citealt{Crenshaw2003}) as the gas intercepts the light from the background quasar. The AGN UV absorption lines are classified according to their widths: narrow absorption lines (hereafter NALs; FWHM $\le$ 500 km s$^{-1}$), broad absorption lines (BALs; FWHM $\ge$ 2,000 km s$^{-1}$) and intermediate class between them (mini-BAL). The relation between these two types of absorption lines has not been understood yet, however one explanation is that these features correspond to different inclination angles (\citealt{Elvis2000}; \citealt{Ganguly2001}). 

Recently, \cite{Rodriguez2020} (RH20 hereafter) presented a survey of UFOs at 2$\le$ z $\le$4.7, probed mainly by CIV and NV absorptions in luminous QSOs (with bolometric luminosity Log($L\rm_{Bol}$/\ergs) = 46.2-47.7, measured for 21 sources) with speeds between 0.1\textit{c} and 0.2\textit{c}, similar to the UFOs typically observed in the X-ray spectra of AGN (\citealt{Tombesi2010}, \citealt{Gofford2013}). They identified 40 UFOs from a starting sample of 6760 BOSS QSOs, finding that 10/40 show extreme velocities (\vufo $\ge$ 50,000 km s$^{-1}$) and six of the 40 QSOs show at least two absorption troughs with velocities larger than 30,000 km s$^{-1}$ up to $\sim$60,000 km s$^{-1}$, and just two of them exhibit three BAL systems. Such large velocities imply large kinetic energy rates, as \ekin $\sim$  \mdot$\times$  \vufo$^{2} \propto$ \vufo$^3$ (since mass outflow rate \mdot$\propto$ \vufo$\times$\rufo$^2$), and UFOs usually exhibit an \ekin\ as large as 10\% of \lbol\ (\citealt{Tombesi2012}, \citealt{Gofford2015}). 
UFOs therefore gained immediate attention as a key mechanism for injecting into the surrounding ISM an amount of energy large enough to significantly affects the host galaxy evolution. The \ekin\ of these winds depends on \vufo, column density (\NH) and distances \rufo\ from the central SMBH. However, accurately constraining these quantities is difficult and typically requires time-consuming, multi-epoch spectroscopy. Although the number of these systems is still limited, observing a variable UFO give us the unique chance of obtaining crucial information on the nature of outflows and their potential effect on the host galaxy.  The time variability of BAL profiles is indeed a powerful tool for studying the origin and the evolution of such outflows, and putting constraints on the lifetime and \rufo.

\cite{Bruni2019} identified a hyperluminous AGN in the WISE/SDSS selected Hyperluminous QSO (WISSH) survey, J1538$+$0855 at z$\sim$3.567, with BAL features exhibiting a maximum velocity of $\sim$0.16\textit{c}. This object has been extensively studied at rest-frame UV/optical wavelengths, exhibiting pervasive signs of outflows at all scales from pc up to tens of kpc (\citealt{Vietri2018}, \citealt{Travascio2020}). \citealt{Vietri2018} reported the presence of broad-line (pc-scale) and narrow-line (kpc-scale) region outflows showing similar kinetic powers, possibly revealing the same outflow in two different gas phases. \citealt{Travascio2020} reported the presence of blueshifted broad emission component of the Ly$\alpha$ providing evidence for outflowing gas reaching distance of 20-30 kpc from the QSO and, for the first time, the detection of a bright $\sim$75 kpc wide nebula in the C IV emission line, thereby revealing a metal-enriched component of the CGM around J1538+0855.

In this paper, we present the discovery of a multi-component ultra-fast BAL outflow in J1538+0855 which exhibits high and complex variability across time.
This finding is based on multi-epoch optical spectroscopy from a set of observations described in Sec. \ref{sec:observations}. In Sec. \ref{sec:results} we perform a study of the kinematic and temporal properties of the three separate BAL components; in Sec. \ref{sec:ion_column} we describe the fitting technique used to get column densities of the absorption troughs, and the photoionization and density analysis. In Sec. \ref{sec:physical_parameters} we report the derived distance and provide an estimate of \ekin. We summarize our results and conclude in Section \ref{sec:conclusions}. Throughout this work we assume a $\Lambda$CDM cosmology with H$_0$ = 70 km s$^{-1}$ Mpc$^{-1}$ and $\Omega_{\Lambda}$ = 0.7.

\begin{table*}
\centering
\begin{threeparttable}

\caption{Journal of rest-frame UV observations.}\label{table:1}             
\centering                          
\begin{tabular}{cccccc}        
\hline\hline                 
Observation ID &  Instrument & R & $\lambda\rm_{range}$ & Observation date & Seeing  \\    
(1) & (2) &(3) &(4) &(5) & (6) \\
\hline                        

SDSS 2006 &  SDSS\textsuperscript{*} &2000 & 3800-9200  & 2006-05-04 &1.4  \\      
SDSS 2012 &   BOSS\textsuperscript{*} &2000 & 3650-10400 & 2012-04-16 & 1.4 \\      
MUSE 2017&   VLT/MUSE & 1750 & 4800-9300  & 2017-07-26 & 0.9\\
MODS  2018&  LBT/MODS & 2000 & 5000-10000  & 2018-06-28 & 0.9\\
FOCAS 2021&  Subaru/FOCAS & 2500 & 3700-6000  & 2021-06-18 & 0.7\\
\hline                                   

\end{tabular}
 \begin{tablenotes}[para,flushleft]
 \item {\bf{Notes}}. (1) Identification name adopted throughout the paper (2) Spectrograph, (3) Resolution, (4) Wavelength range of the instrument in \AA, (5) Dates and (6) seeing of the optical observations.
 
 (*) Plate, MJD, and fiber identifying the SDSS spectrum is 1724-53859-376 and for the BOSS spectrum is 5206-56033-202.
\end{tablenotes}
\end{threeparttable}

\end{table*}
\section{Observations}\label{sec:observations}

In this paper we present the analysis of the following spectroscopic data:

\begin{itemize}
\item[\textbullet]  \textit{SDSS} and \textit{BOSS}. The SDSS observations were carried out as part of SDSS DR10 and DR14 releases (\citealt{Gunn2006}). The observation of SDSS spectrum was performed on May 4, 2006 and cover a broad wavelength range spanning from 3800 to 9200 \AA\ at a spectral resolution of R $\sim$ 2000. The BOSS observation was performed on April 16, 2012 and taken with the same telescope as SDSS ones but with a different spectrograph covering a wavelength range from 3650 and 10400 \AA\ (R $\sim$ 2000).

\item[\textbullet] \textit{VLT/MUSE}. J1538+0855 was observed with MUSE on July 26, 2017, as part of the ESO program ID 099.A-0316(A) (PI F. Fiore). The observation consists of four exposures of 1020s each. The average seeing was $\sim$0.9 arcsec. We refer to \cite{Travascio2020} for a detailed description of the data reduction.

\item[\textbullet] \textit{LBT/MODS}. The observation with the MODS spectrograph on the LBT was performed on June 28, 2018. We obtained four exposures of 675s using the red G670L grating with a 1 arcsec slit width and the seeing was $\sim$0.9 arcsec during the spectra acquisition. The spectra of the star BD+33 2642 were also obtained to flux calibrate and remove the atmospheric absorption, and arc-lamps were used for the wavelength calibration. These spectra were flat-field corrected and wavelength-calibrated by the INAF–LBT Spectroscopic Reduction Center in Milan, where the LBT spectroscopic pipeline was developed. Flux calibration and telluric-absorption correction were performed with own python routines.

\item[\textbullet] \textit{Subaru/FOCAS}. We conducted spectroscopic observation with Subaru/FOCAS (PI. T. Misawa) on June 18, 2021, using VPH650 grating with a slit width of 0.4 arcsec. The total integration time was 3600 s. The typical seeing was 0.7 arcsec during the observation. The spectroscopic standard star Hz44 was also observed. We reduced the FOCAS data in a standard manner with the software IRAF. Wavelength calibration was performed using a Th-Ar lamp.

\end{itemize}

Table \ref{table:1} lists a summary of the optical spectroscopic observations of J1538+0855 analyzed in this paper.

\section{Results}\label{sec:results}
\subsection{BAL detection}\label{sec:BAL_det}

We first normalize each spectra to the continuum, to correctly identify the absorption features. To this end, we fit the continuum with a power-law function and emission lines with Gaussian components for Ly$\alpha$, NV, SiIV, CIV, HeII and CIII], masking the regions showing absorption troughs, which can affect the continuum estimation. The model fitting is performed in the spectral range 1210-2000 \AA\ excluding the spectral region 1570-1631 \AA, i.e. the unidentified emission feature (\citealt{Nagao2006}) and  1687-1833 \AA\ where there are heavily blended emission lines as NIV$\lambda$1719, AlII$\lambda$1722, NIII]$\lambda$1750 and FeII multiplets (\citealt{Nagao2006}). The best-fit model is derived through the $\chi^2$ minimization. We also follow the prescription presented in RH20 as an alternative method to normalize the quasar spectra (see Appendix \ref{sec:appendix1} for more details). 
Fig. \ref{fig:rodriguez} shows the best-fit model of the continuum following both procedures. In all but FOCAS observations, our continuum estimate is consistent with that of RH20 by using R$_1$, R$_2$ and R$_4$ regions to define the underlying power-law continuum (see Fig. \ref{fig:rodriguez} and Appendix \ref{sec:appendix1}).

\begin{figure}
   \centering
   \includegraphics[width=9cm]{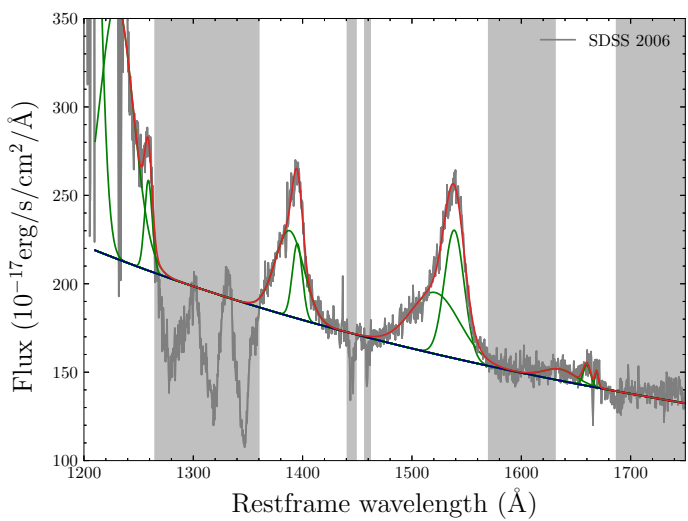}
      \caption{SDSS 2006 spectrum of J1538+0855 (grey), showing our best-fit model (red). We fit a power law (blue) to the spectrum and Gaussian functions for the emission lines (green curves). The grey bands represent the regions masked during the continuum and emission lines fit.}\label{fig:rodriguez}
   \end{figure}

We measure the Balnicity index (BI) for each observation, as a proxy for defining the BAL systems and to provide a measure of the total BAL strength. We use the definition of BI as in RH20, which is slightly different from the original definition introduced by \cite{Weymann1991}, since it assumes the minimum and maximum velocity of a BAL-trough region at 30,000 and 60,000 km s$^{-1}$, respectively, corresponding to the velocity interval of an absorption system associated with CIV line lying between NV and SiIV emission lines.  
To identify and characterize the absorptions features in the velocity space, the spectral regions must contain contiguous absorption that reaches $\ge$ 10 per cent below the continuum across at least 1000 km s$^{-1}$:

\begin{equation}
\rm BI= - \int_{60,000}^{30,000} \Bigg[1-\frac{\textit{f(v)}}{0.9}\Bigg]\textit{C}  \,dv 	\label{eq:1}    
\end{equation}

where \textit{f(v)} is the continuum-normalized flux, which is divided by 0.9 to avoid shallow absorptions, where the square bracket is positive if the absorption fluxes are below the 90\% level of the normalized flux. The \textit{C} factor can assume two values, i.e. 1 when the square bracket value is continuously positive over a velocity interval of our choice and zero otherwise. 

 \begin{figure*}
   \centering
   \includegraphics[width=16cm,height=10cm]{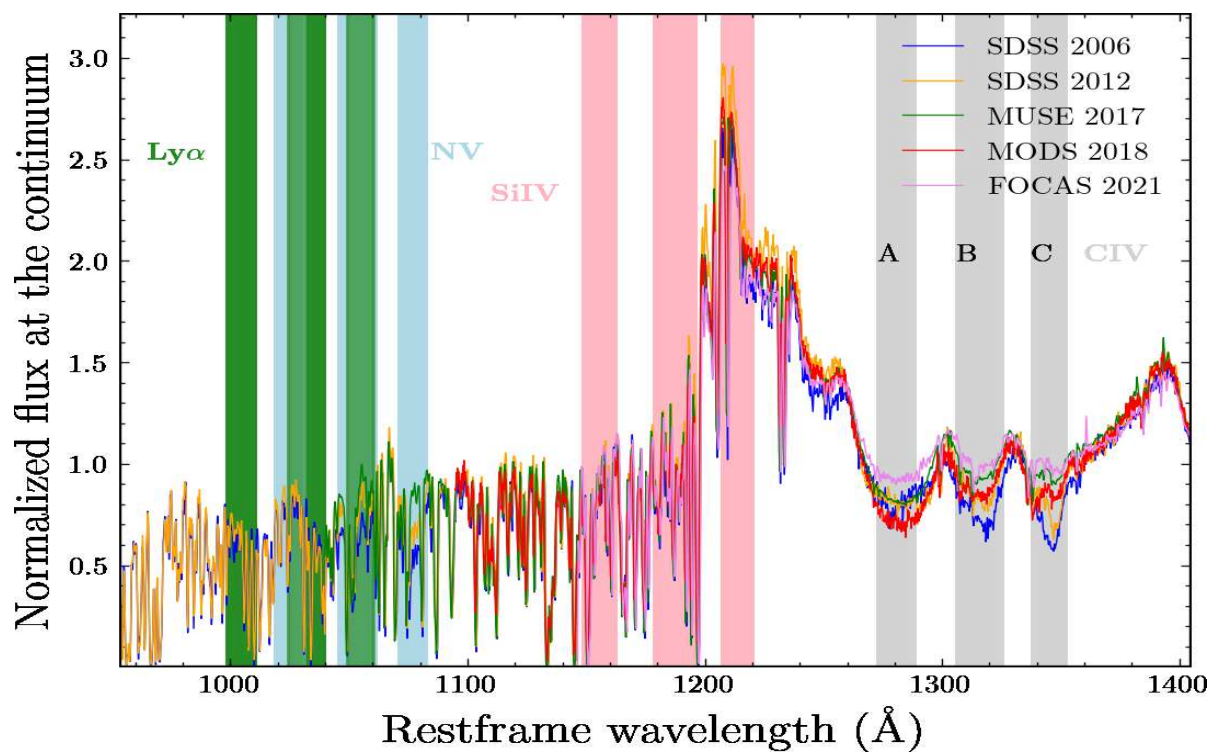}
      \caption{Comparison of the  optical spectra of J1538+0855 taken at different epochs (see Table \ref{table:1}, with  the observed locations of CIV absorption troughs A, B and C marked as grey shaded regions. The expected positions of the corresponding SiIV (pink), NV (cyan), Ly$\alpha$ (green) and OVI (coral) absorption are also shown as shaded regions.}\label{fig:other_BAL}
   \end{figure*}
\begin{figure}
   \centering
   \includegraphics[width=9cm]{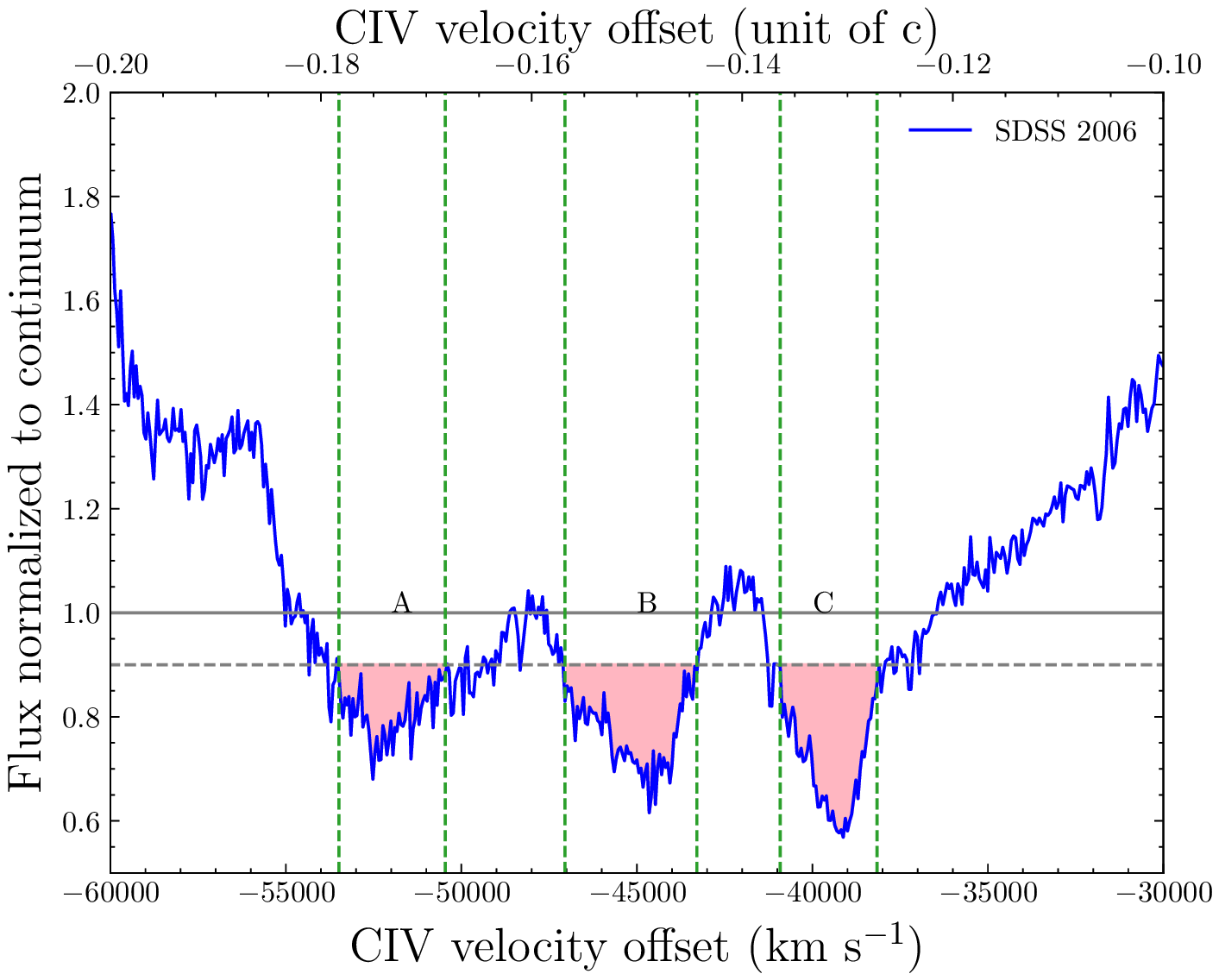}

      \caption{The SDSS 2006 continuum-normalized spectrum of J1538+0855, showing highly-blueshited (> 0.1\textit{c}) BAL signatures (A, B and C) associated with CIV absorption, blue-wards of the SiIV emission line. In light pink are indicated absorptions below 90\% of the continuum. Green dashed lines represent the minimum and maximum velocity estimated for the BAL outflow in each trough, respectively. Horizontal dashed line represents the 90\% level of the  continuum-normalized flux (solid line).}\label{fig:velocity}
   \end{figure}

We identify a three component absorption system, namely A, B and C (see Fig. \ref{fig:other_BAL}), in the range $\sim$1240-1390 \AA, which we associate with the CIV line. Usually, CIV BALs are found between SiIV and CIV emission lines, and absorption appearing between Ly$\alpha$ and SiIV, as in our case, can be caused by SiIV absorption. In the case of J1538+0855 we can rule out the hypothesis of absorptions due to SiIV, due to the lack of CIV absorptions, at similar velocities. Indeed no cases are reported in the literature with SiIV absorption without the corresponding CIV absorption (\citealt{FilizAk2013}; \citealt{Bruni2019}; RH20 for a detailed discussion). Given their extremely high velocities (>0.1\textit{c}, see Fig. \ref{fig:velocity}), these CIV absorbers can be classified as UFOs. 

\begin{figure}
   \centering
   \includegraphics[width=9cm]{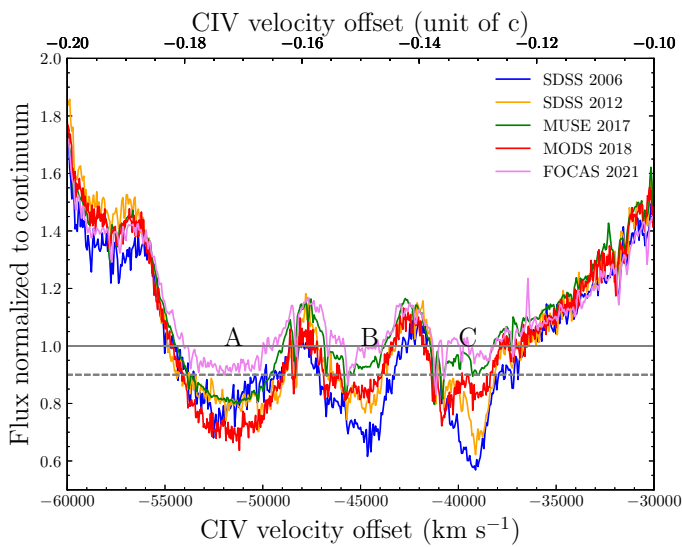}

      \caption{Normalized spectra of J1538+0855 in the rest-frame 1240-1390 \AA\ region, where the three highly-blueshited CIV BAL are detected. Horizontal dashed line represents the 90\% level of the  continuum-normalized flux (solid line).}\label{fig:norm}
   \end{figure}
   
The absorption troughs show a large variation in profile and strength at different epochs, with a partial fading of the BAL in 2017 (B and C troughs) up to the total disappearance of the BAL system in 2021 (see Fig. \ref{fig:norm}). The A trough has a smooth profile and mostly varies in strength, differently B and C troughs show sub-components undergoing distinct changes. Specifically, the C trough exhibits a sub-component at low velocity which significantly weakens over time, producing a big change in the total profile. This could be likely due to opacity variation in the absorption trough limited to the low velocity gas.

\begin{figure}
   \centering
   \includegraphics[width=9cm]{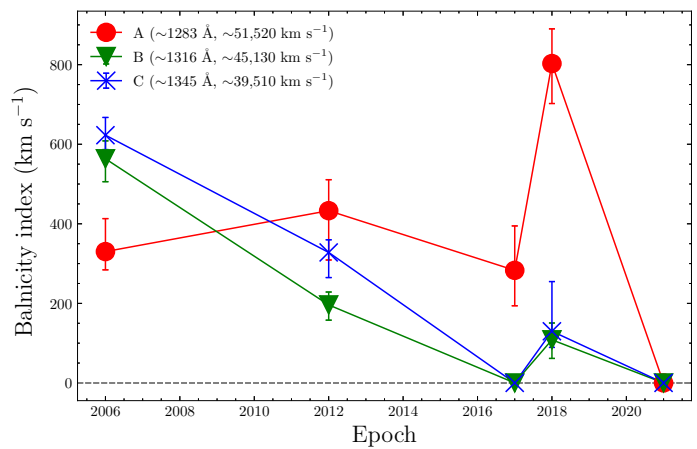}

      \caption{Balnicity index of A, B and C BAL troughs in J1538+0855 at each observing epoch. Grey dashed line indicates the zero level.}\label{fig:BI}
   \end{figure}

The exceptional variability of the three BAL components is shown in Fig. \ref{fig:BI}, with varying BI values, for the deepening or weakening of the troughs, or the disappearance of a specific trough (BI=0 for B and C troughs in 2017 epoch) up to the disappearance of the total BAL system in 2021 epoch (BI=0 for A, B and C troughs). The BAL velocity, estimated from the minimum and maximum wavelengths of the BAL troughs, are in the range $\sim$0.13-0.18\textit{c}, as shown in Fig. \ref{fig:velocity}, with values consistent between all epochs for each trough (see Appendix \ref{sec:appendix2}). 
Table \ref{table:2} lists the BI and velocities values of the three components measured from all the spectra. We note that FOCAS spectrum  shows weaker CIV absorption features, which do not satisfy the BI criterion.

We search for SiIV, NV, OVI and Ly$\alpha$ absorption features with the same \vufo\ as found for the the CIV troughs.  Fig. \ref{fig:other_BAL} shows the normalized spectra with marked as grey shaded regions the observed locations of CIV absorption troughs A, B and C, along with the expected positions of the corresponding SiIV, NV, Ly$\alpha$. We note that MUSE, MODS and FOCAS data do not cover the spectral region associated with possible ultra-fast NV and Ly$\alpha$ BALs. 
We are able to recover SiIV, NV and Ly$\alpha$ BAL systems in the SDSS spectra (see Sec. \ref{sec:ion_column}  for further details).
 
Furthermore, a comparison between epochs reveal that at the velocity of the A, B and C CIV troughs, the flux variation computed for different ions appears to modulate similarly to the corresponding CIV troughs (see Fig. \ref{fig:spectra_ratio}). The implications of this behaviour are discussed in Sec. \ref{sec:variability}.


 
  
%
 
%
 
\subsection{Variability}\label{sec:variability}

We measure the strength of the absorption troughs A, B and C in all the spectra, to study their variability in detail. 
We calculate for each epoch the absorption strength, \textit{A$_S$}, as described in \cite{Capellupo2013}, which is defined as the fraction of the normalized continuum flux removed by absorption (0$\le$ \textit{A$_S$}$ \le$1) within the velocity interval that varied between two consecutive epochs (the higher the absorption feature, the higher \textit{A$_S$}). 
A single value of \textit{A$_S$} is adopted for each trough, based on the average flux within the velocity interval that varies. The latter must be at least 1200 km s$^{-1}$ in width and the flux difference in this region must be at least 4$\sigma$ (see eq. 1 in \citealt{Capellupo2012}) to be included as a varying region (Fig. \ref{fig:variability} highlights the varying region found between the first two epochs, i.e. 2006 and 2012, see Appendix \ref{sec:appendix3} for the other epochs). 

The value of \textit{A$_S$} and $\Delta$\textit{A$_S$} give a direct measurement of the strength of the BAL and the change in absorption strength between two epochs, respectively, in only the varying spectral region. As reported in Fig. \ref{fig:variability}, we find that CIV A trough shows a deepening (i.e. \textit{A$_S$} is higher than the previous epoch) from epochs 2006 to 2012 at the low velocity end of the absorption on timescale of \trest=1.30 year, with no significant variation of the trough at higher velocity. Fig. \ref{fig:D1_A} in Appendix \ref{sec:appendix3} shows the varying regions for the A troughs between epochs (i) 2012-2017, where it gets shallower in a similar velocity range as that of 2006-2012, (ii) 2017-2018, where it gets deeper over the total velocity interval of the trough, and (iii) 2018-2021, where the BAL disappears (see Appendix \ref{sec:appendix3}). 

\begin{figure}
   \centering
   \includegraphics[width=9cm]{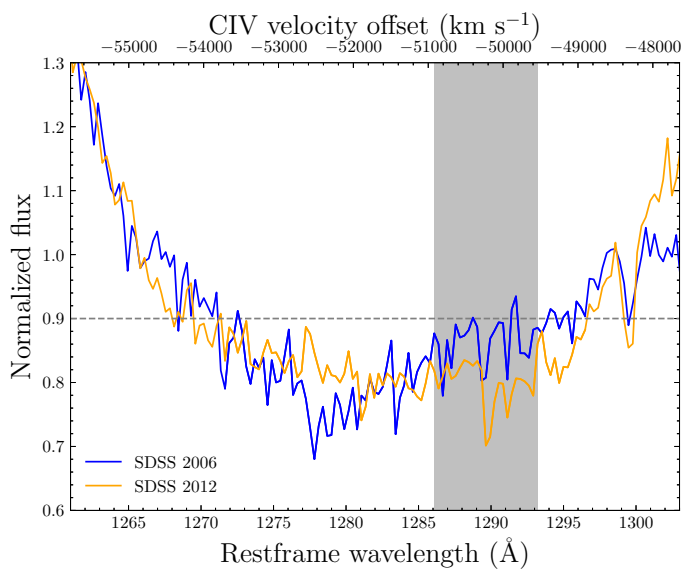}
 \caption{Comparison of the spectra at the location of the CIV A trough, between SDSS 2006 and SDSS 2012 epochs. Shaded region marks the interval of velocity that varied between the two epochs, within which the parameter \textit{A$_s$} (the fraction of the normalized continuum flux removed by absorption) was derived, for each epoch. Dashed line represents the 90\% level of the continuum-normalized flux.}\label{fig:variability}
   \end{figure}

 As for B and C troughs, they change in concert, in spite of their different outflow velocity (i.e., they are caused by absorbers with similar physical condition): they gradually disappear from 2006 to 2017, after that they get deeper in 2018 and disappear again in 2021 (see Fig. \ref{fig:D1_B} and \ref{fig:D1_C} in Appendix \ref{sec:appendix3}).

   \begin{figure}
   \centering
   \includegraphics[width=9cm]{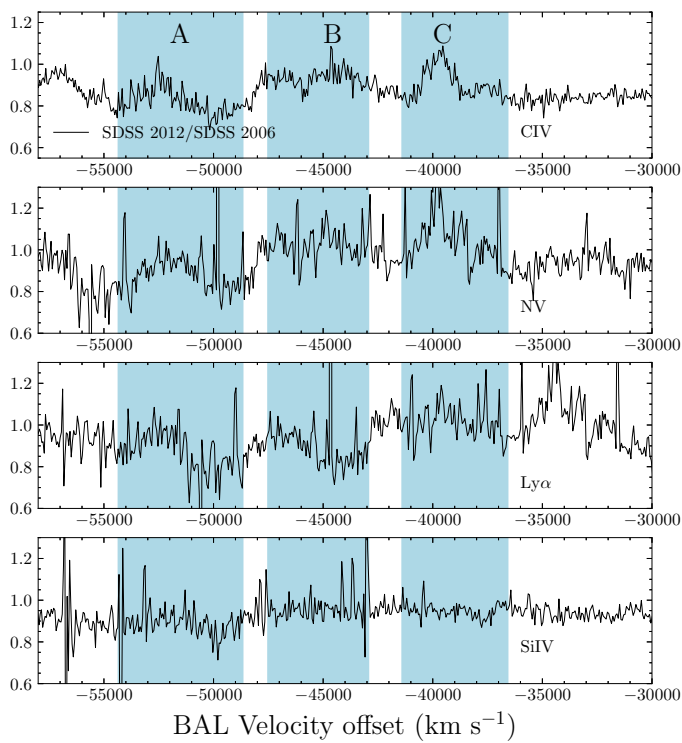}
      \caption{The SDSS 2012 and SDSS 2006 ratio spectra for CIV (top panel), NV (second panel), Ly$\alpha$ (third panel) and SiIV (bottom panel) absorption regions. The light-blue bands highlight the common variability spectral range of A, B and C BAL absorptions features in the velocity space.}\label{fig:spectra_ratio}
   \end{figure}

Two different scenarios can explain the observed variability of the different components of the BAL UFO, i.e. the motion of the gas across our line of sight and changes in ionization. The fact that the variability behavior of A trough is different from that of B and C components clearly suggests a different origin among them. Fig. \ref{fig:spectra_ratio} shows the ratio of SDSS 2012 to SDSS 2006 spectrum, which are the only two spectra covering the position of the corresponding ultra-fast NV and Lya BALs. On the one hand, looking at the variation in the CIV and in the spectral regions corresponding to other ions with a different ionization state such as SiIV, Ly$\alpha$ and NV, we find that for troughs B and C, SiIV is almost stable, while CIV, NV and Ly$\alpha$ are variable, which is inconsistent with the gas motion scenario. On the other hand, for trough A, the variability of SiIV matches that observed for CIV, NV and Ly$\alpha$ transitions. These findings are confirmed by the Spearman correlation test run over the flux ratio between the SDSS 2006 and 2012 epochs in the spectral region associated with the CIV BAL and that of NV, Ly$\alpha$ and SiIV assuming the same velocity (see Table \ref{tab:3}). This result supports a gas motion scenario for trough A, at least for these two epochs and a change in ionization as the cause of the observed variability for the B and C troughs. We compare the BI and the continuum luminosity at 1450 \AA, L$_{1450}$, for troughs B and C excluding the two epochs with BI=0, finding a strong correlation (p-value$\ll$1$\times$10$^{-4}$) (\citealt{Trevese2013}), while BI and L$_{1450}$ do not correlate for trough A (excluding the last epoch with BI=0, p-value=0.6), suggesting the gas motion as the origin of its variability (\citealt{Gibson2008}).

 \begin{table}
 \begin{threeparttable}

\caption{Results of the Spearman's correlation test of the ratio between SDSS 2006 and 2012 spectra.}\label{tab:3}              
\centering                          
\begin{tabular}{c c c c}        
\hline\hline                 
BAL component & Ions & $\rho$ & P-value \\
(1) & (2) & (3) & (4)\\
\hline                        

A  & CIV-NV &  0.38 & 1.2$\times$10$^{-4}$ \\
 & CIV-Ly$\alpha$ &  0.44 & 5.3$\times$10$^{-6}$ \\
 & CIV-SiIV &  0.29 & 3.5$\times$10$^{-3}$ \\

B & CIV-NV &  0.12 & 0.30 \\
 & CIV-Ly$\alpha$ &  -0.21 & 0.07 \\
 & CIV-SiIV& 0.18 &  0.11 \\

C & CIV-NV &  0.58 &  1.0$\times$10$^{-8}$ \\
 & CIV-Ly$\alpha$ &  0.39 &  4.1$\times$10$^{-4}$ \\
 & CIV-SiIV &  -0.14 & 0.21\\

\hline                                   
\end{tabular}
 \begin{tablenotes}[para,flushleft]
 \item {\bf{Notes}}. (1) Component of the BAL UFO, (2) Spectral ratio comparison between different ions, (3) Spearman rank and (4) Significance level. 
\end{tablenotes}
\end{threeparttable}

\end{table}

These findings allow us to adopt a hybrid scenario in which both gas motion and change in ionization are responsible for BAL variability in J1538+0855.

\section{Ionization and total column density}\label{sec:ion_column}

 \begin{figure}
  \centering
   \includegraphics[width=9cm]{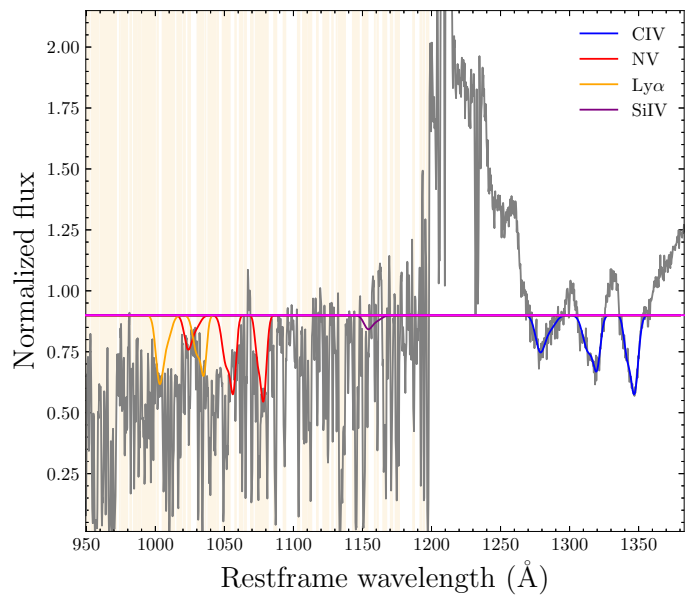}
      \caption{Continuum-normalized SDSS 2006 spectrum showing from right to left the best-fit model of the UFO BALs troughs for CIV (blue), SiIV (purple), NV (red) and Ly$\alpha$ (gold) ions. The spectrum includes many lines of the Ly$\alpha$ forest that we mask (vertical gold lines) during the fitting procedure. The continuum level is represented as magenta horizontal line. }\label{fig:ions_trough}
   \end{figure}
   
\begin{figure}
  \centering
   \includegraphics[width=8cm]{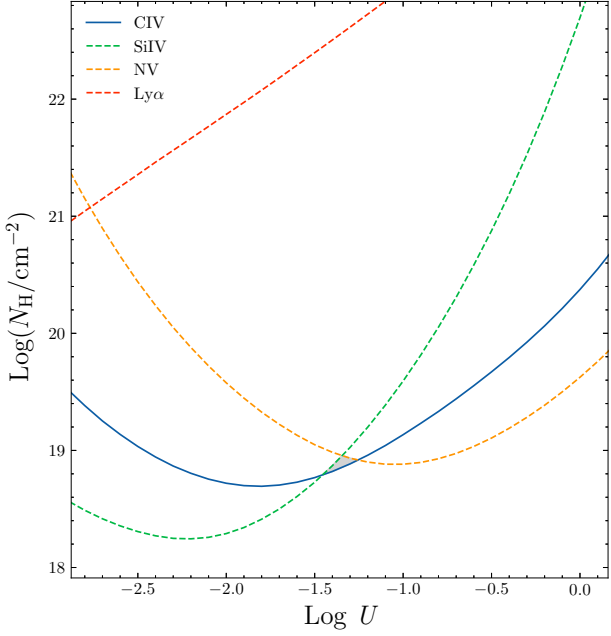}
   \caption{Theoretical values of \textit{U} and computed $N\rm_H$ by assuming the $N\rm_{ion}$ lower limit for CIV (blue solid curve), and upper limits for NV (orange dashed curve), SiIV (green dashed curve) and Ly$\alpha$ (red dashed curve) ions for the A component of the UFO BAL in the SDSS 2012 spectrum of J1548$+$0855.}\label{fig:curves}
   \end{figure}

A careful analysis of the SDSS spectra in the region on the shortward side of the Ly$\alpha$ emission allows us to recover BAL features corresponding to NV, Ly$\alpha$ and SiIV lines at the same velocity as CIV BAL troughs (see Fig. \ref{fig:ions_trough}). To derive their ionic column densities $N\rm_{ion}$, we model the normalized spectrum $I\rm(\lambda)$=$F\rm_{obs}(\lambda)$/$F\rm_c$ where $F\rm_{obs}$ is the observed flux and $F\rm_c$ the continuum flux. We adopt the apparent optical depth (AOD, \citealt{Savage1991}) method to measure $N\rm_{ion}$\footnote{We treat unresolved doublets as single lines, with a summed oscillator strength and weighted average wavelength (e.g. \citealt{Hamann2018}).}, relying on the $\tau(\lambda)\equiv -ln(I(\lambda)$) relation between the intensity and the optical depth. The AOD method is used to find lower limits on $N\rm_{ion}$ for singlets and contaminated doublets, or upper limits on $N\rm_{ion}$ in case there are no detection of absorption troughs.\\ 
We first fit the CIV troughs by assuming the following Gaussian optical depth profiles (in the velocity frame):
\begin{equation}
    \tau({\rm v})=\tau_0 e^{-\Delta {\rm v}^2/b^2}
\end{equation}

where $\tau_0$ is the line-center optical depth, $\Delta$v is the velocity shift from the line center, \textit{b} is the Doppler parameter. We adopt two Gaussian components to account for the asymmetric profiles of each trough. 
We then fit the absorption from NV, Ly$\alpha$ and SiIV ions at the velocities measured for the CIV UFO by adopting the following procedure:
\begin{itemize}
    \item $\Delta {\rm v}$ and $b$ for each component of NV, Ly$\alpha$ and SiIV absorption lines are fixed to match the best-fit values of the CIV troughs;
    \item the ratio of the $\tau_0$ of the two Gaussian components describing each trough are fixed to that of the corresponding CIV trough;
    \item the narrow absorption lines from the Ly$\alpha$ forest are masked.
\end{itemize}
This avoids blending with the Ly$\alpha$ forest features and allows to place upper limits on $N\rm_{ion}$.\\
By integrating the best-fit optical depth profiles over the troughs:
\begin{equation}
N{\rm_{ion}}=\frac{3.7679 \times 10^{14}}{\lambda f} \int \rm \tau(v)dv 
\end{equation}

where $\lambda$ is the laboratory wavelength and \textit{f} is the oscillator strength of the considered species. We use the numerical code Cloudy 17.02 version (\citealt{Ferland1998},\citealt{Ferland2017}) to compute the fraction of ionic species by varying the ionization parameter \textit{U} and assuming a gas in photoionization equilibrium, solar abundances (\citealt{Lodders2003}) and the broad-band spectral energy distribution of J1538+0855 (Saccheo et al. in prep). Fig. \ref{fig:curves} shows the theoretical curves of \textit{U} and measured  \textit{$N\rm_H$} by assuming the column density $N\rm_{ion}$ derived for CIV, NV, SiIV and Ly$\alpha$ ions for the A component of the UFO BAL in J1548+0855. The grey shaded region in Fig. \ref{fig:curves} represents the range of $N\rm_H$ and \textit{U} values that are consistent with both upper limits on SiIV and NV and the lower limit of the CIV, respectively. We adopt the mean value of these ranges as the best estimate for the ionization parameter and total column density, i.e. \textit{U}$\approx$0.04  and $N\rm_H$ $\approx$ 1.2 $\times$10$^{19}$ \cmdue, respectively, taking into account relativistic effect (\citealt{Luminari2020}). Following the same method, we also derive \textit{U}$\approx$0.02 and $N\rm_H$ $\approx$1.6$\times$10$^{19}$ \cmdue\ both for B and C troughs.

\section{Physical parameters of the BAL UFO in J1538+0855}\label{sec:physical_parameters}
\subsection{Estimate of the distance of the ultra-fast outflows}\label{sec:distance}
\subsubsection{Component A of the UFO}

The gas transverse motion scenario allows to place an upper limit on the distance of the absorbing gas responsible for the A component of the UFO from the continuum source. Following \cite{Capellupo2013} we estimate the size of the continuum region at a particular wavelength, D$_{\lambda}$. For estimating D$_{\lambda}$, we consider a standard thin accretion disc that emits as a blackbody at an effective emitting wavelength $\lambda$ and use Wien’s displacement law ($kT=hc/\lambda$, with $h$ the Planck constant, $k$ the Boltzmann constant and \textit{c} the speed of light) to translate the temperature to the wavelength corresponding to maximum blackbody emission to estimate the radius of the disc at that wavelength (\citealt{Peterson1997}). Following \citealt{Morgan2010}, the size of the accretion disk can be rewritten as:
\begin{equation}
R_{\lambda\rm_{max}}=9.7 \times 10^{15} \Bigg(\frac{\lambda\rm_{max}}{\rm\mu m}\Bigg)^{4/3}\Bigg(\rm\frac{\textit{M}_{BH}}{10^9 M_{\odot}}\Bigg)^{2/3} \Bigg(\rm\frac{\textit{L}_{Bol}}{\eta \textit{L}_{Edd}}\Bigg)^{1/3} cm  
\end{equation} 

where \Mbh\ is the BH mass, \lbol/\ledd(=\Edd) the Eddington ratio, $\eta$ the radiative efficiency, and $\lambda\rm_{max}$ the maximum blackbody emission wavelength. We used \Mbh=5.5 $\times$ 10$^9$ M$_{\odot}$ and \Edd=1, as found by \cite{Vietri2018}. Adopting a standard $\eta$=0.10 and $\lambda\rm_{max}$=1300 \AA\ based on the location of the CIV BAL troughs in J1538+0855, we find D$_{1300}$=2$\times R_{1300}$=0.003 pc.

We then derive the absorption strength \textit{A$_S$} and measure its variations ($\Delta$\textit{A$_S$}) between the two epochs consistent with gas motion scenario (2006-2012 yr). We calculate the crossing speed of the outflow component, as indicated by \citealt{Capellupo2013}. Specifically, we assume their crossing disk model and that the A trough in J1538+0855 is always saturated during our monitoring campaign: a circular disc crosses a larger circular continuum source along a path through the centre of the background circle. In this case the distance travelled by the A absorber is \textit{d}=$\sqrt{\Delta A_S}$D$_{1300}$ and considering a \trest=1.30 yr between SDSS 2006 and 2012 for which a transverse gas motion scenario is favoured, the derived transverse velocities is $v\rm_{cross}$=\textit{d}/\trest= 640 km s$^{-1}$ to cross the continuum source. 
Assuming that the crossing speed is equal to the Keplerian rotation speed around the BH, we constrain the location of the A component of the BAL UFO in J1538$+$0855 at \routA$\le$58 pc.

\subsubsection{Components B and C of the UFO}
Change in the ionizing flux scenario sets an upper limit on the distance between the clouds responsible for B and C troughs and the continuum source. In this case, by assuming that the gas is in photoionization equilibrium and following \cite{Narayanan2004}, the ionization parameter is given by:
\begin{equation}
    U= \frac{1}{4\pi R{\rm_{ufo}^2} n{\rm{_e}} c} \int_{0}^{912 \AA} \frac{\lambda L_{\lambda}}{hc} \,d\lambda 
\end{equation}

with $\lambda L_{\lambda}$ the spectral energy distribution of J1538+0855 and  $n\rm_e$ the hydrogen number density, which is function of the recombination time $t\rm_{rec}$ as follows:

\begin{equation}
    n{\rm_e}\sim\frac{1}{{\alpha\rm_{rec}} {t\rm_{rec}}}
\end{equation}

where $\alpha \rm_{rec}$=2.8$\times 10^{-12}$ \cmtre \suno\ (\citealt{Arnaud1985}) is the recombination rate coefficient for CIV $\rightarrow$ CIII. By assuming a recombination time equal to  \trest=1.30 yr between SDSS 2006 and 2012, it is therefore possible to derive a lower limit on $n\rm_e$ $\ge$8,700 \cmtre.
This results into an upper limit on \rufo\ for the B and C troughs at \routBC $\le$2.7 kpc, adopting \textit{U}$\approx$0.02 as found from Cloudy photoionization simulations (see Sect. \ref{sec:ion_column}).

\subsection{Kinetic power of the ultra-fast outflows}\label{sec:kinetic}
We also provide an estimate of the kinetic power for each component of the BAL UFO following \cite{Hamann2019}. We estimate the kinetic energy $E\rm_{K,ufo}$, by assuming an expanding shell at a certain velocity (\vufo) with a global covering factor $Q$\footnote{The outflow clumpiness is not taken into account, i.e. we assume a volume filling factor of unity, therefore \ekin\ is to be considered as an upper limit.} (the fraction of 4$\pi$ steradians covered by the outflow as seen from the central quasar) and at a radial distance, \rufo:

{\small{
\begin{equation}
{E}\rm_{K,ufo}= 4.8 \times 10^{53} \Bigg(\frac{{\it{Q}}}{0.15}\Bigg) \Bigg(\frac{\textit{N}\rm_H}{5\times 10^{22}\ cm^{-2}}\Bigg) \Bigg(\frac{\textit{R}_{ufo}}{1\ pc}\Bigg)^2\Bigg(\frac{\textit{\rm{v}}_{ufo}}{8000\ km\ s^{-1}}\Bigg)^2 erg.\label{eq:ekin} 
\end{equation}}}

For the component A we derive ${E}\rm_{K,ufo}$ by using \rufo=\routA$\le$58 pc, \vufo=v$\rm_{max}$=53,700 km/s,\footnote{By assuming \vufo=v$_{50}$, the velocity at 50 per cent of the line flux, as done for X-ray UFOs, \ekin\ decreases by a factor of $\sim$13\%.} i.e. the velocity estimated from the maximum wavelengths, representative of the bulk velocity of the outflowing gas (see Fig. \ref{fig:velocity} and Table \ref{table:2}), {\it{Q}} = 0.15 (based on the incidence of CIV BALs in SDSS QSOs, e.g. \citealt{Gibson2009}, \citealt{Hamann2019}) and \NH = 1.2$\times 10^{19}$ \cmdue\ as derived from photoionization models (see Sec. \ref{sec:ion_column}). Dividing ${E}\rm_{K,ufo}$ by a characteristic flow time given by \rufo/\vmax, we find a kinetic power of \ekin$\le$ 5.2$\times10^{44}$ \ergs.  
Similarly, by assuming \rufo=\routBC\ and \NH = 1.6$\times 10^{19}$ \cmdue\ we find \ekin $\le$2.1$\times$10$^{46}$ \ergs\ and \ekin $\le$1.4$\times$10$^{46}$ \ergs, for the trough B and C, respectively.

\begin{figure}
   \centering
   \includegraphics[width=9cm]{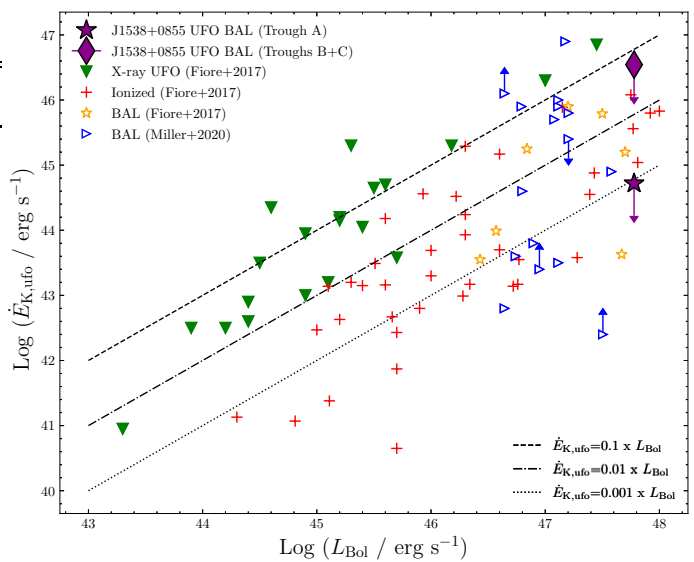}
      \caption{Kinetic power as a function of bolometric luminosity for CIV BAL J1538+0855 (red and magenta stars) and a compilation of AGN showing different types of outflows from \cite{Fiore2017}, i.e. ionized [OIII] (red crosses), BAL outflows (orange stars) and X-ray UFOs (green triangles), and from \cite{Miller2020}, BAL outflow (blue open triangles).}\label{fig:energetics}
   \end{figure}

In Fig. \ref{fig:energetics} we report the \ekin\ values for A, B and C components in J1538$+$0855 as a function of the \lbol\ compared to the values reported for a large collection of outflows by \cite{Fiore2017} and \cite{Miller2020}. The upper limit constraints on the distance of the different components of the UFO BAL imply to \ekin\ values consistent with $\sim$0.1-10\% of the \lbol\ of J1538$+$0855.

 \begin{table}
 \begin{threeparttable}

\caption{Properties of the multi-component BAL in J1538$+$0855}\label{table:2}              
\centering                          
\begin{tabular}{c c c c c c}        
\hline\hline                 
BAL component & Epoch & BI & \vmin & \vmax  & $\Delta$t$\rm_{rest}$\\
(1) & (2) & (3) & (4) & (5) & (6)\\
\hline                        

A & 2006 &  330$_{+80}^{-50}$ & 50460 & 53490& - \\
  & 2012 &  430$_{+80}^{-120}$ & 48900 & 53700& 1.30 \\
  & 2017 &  280$_{+110}^{-90}$ &49660 & 53580 & 1.16\\
  & 2018 & 800$_{+90}^{-100}$ & 49000 & 54010 & 0.20 \\
  & 2021 & - & - & -& 0.65\\

B & 2006 &   560$_{+40}^{-60}$ & 43290  &47050 &  - \\
  & 2012 &   200$_{+30}^{-40}$ & 43940 & 46450 &  1.30 \\
  & 2017 &  -  & - & - & 1.16\\
  & 2018 &  110$_{+40}^{-50}$ & 43950   & 46010 & 0.20 \\
  & 2021 &  -  & - & - & 0.65 \\

C & 2006 &  620$_{+40}^{-40}$ & 38160 & 40920 & -\\
  & 2012 &  330$_{+30}^{-60}$ & 38160 & 40920 & 1.30  \\
  & 2017 &  - & - & - & 1.16 \\
  & 2018 & 130$_{+120}^{-40}$  & 39840 & 41360 & 0.20\\
  & 2021 & - &- &- & 0.65  \\
\hline                                   
\end{tabular}
 \begin{tablenotes}[para,flushleft]
 \item {\bf{Notes}}. (1) Component of the BAL UFO, (2) observation epoch, (3) Balnicity index, (4) minimum and (5) maximum velocity of the BAL trough in km/s and (6) rest-frame time in year elapsed since the previous observation. 
\end{tablenotes}
\end{threeparttable}

\end{table}

\section{Summary and Conclusions}\label{sec:conclusions}

We have presented the analysis of extremely high-velocity and highly variable CIV BAL troughs discovered in the luminous QSO J1538$+$0855 at z$\sim$3.6, based on spectroscopic observations taken at five different epochs spanning about 17 years in the observed frame (see Table \ref{table:1}).

We report on the detection of a three-component BAL UFO ($\sim$40,000-54,000 km s$^{-1}$), showing variability both in strength and shape, disappearing and reappearing at different epochs. 
The number of similar UFOs detected so far is still limited (only 2 out of 40 BAL UFOs in RH20 show three distinct components). The study of the evolution of the BAL features in this unique system gives us the opportunity of shedding light on the physical properties of the BAL UFO and its potential effect on the host galaxy.

By looking at the presence of NV, OVI and SiIV absorptions at similar speeds of CIV BALs, we found that the troughs changed in a similar manner and the variability trends are not the same for all components i.e., the components B and C varied in concert, while the A component exhibits a different behaviour.
These findings allow us to support a hybrid scenario as the origin of the variability, i.e. changes in ionization state for B and C troughs and transverse motion of the absorbing material responsible for the A component across our line of sight between 2006 and 2012. 
The disappearance of B and C troughs observed during the 2017 epoch, and the total disappearance of all the three BAL components observed in 2021, could be the result of the combination of such scenarios. On the one hand, for B and C components a sufficient increase in the ionization potential and/or a reduction in the amount of shielding material between the radiation source and the outflow could cause the coordinated disappearance of the troughs, leading to reduced ionic column densities (e.g. \citealt{Decicco2018}). On the other hand, the BAL material responsible for the A component could just be moved out from our line of sight causing its disappearance, but still existing (e.g. \citealt{Capellupo2012}, \citealt{Filizak2012}, \citealt{Decicco2018}). The transverse motion scenario responsible for the variability observed during 2006 and 2012 epochs implies an upper limit of \routA$\le$58 pc for the outflowing gas of the A trough if Keplerian-dominated tangential motion is assumed. Variations of the A trough have been confirmed by subsequent observations but the lack of spectroscopic data at $\lambda \le$1200 \AA\ does not allow to pinpoint their origin. 
By assuming that the variations observed for CIV A component are due to gas motion, in particular during the epochs 2017 and 2018, corresponding to the shortest observed variability time (\trest=0.20 yr), the transverse velocity is  v$\rm_{cross}$= 6,500 km s$^{-1}$ leading to a minimum upper limit estimate of the distance \routminA $\le$0.6 pc.
This absorber would be placed at a distance consistent with the CIV broad line region radius of J1538+0855 ($R\rm^{CIV}_{BLR}$=0.4 pc, \citealt{Lira2018}). This would be in agreement with the expectation of some popular models of BAL outflows (\citealt{Murray1995}, \citealt{Proga2000}), which predict that they are co-spatial (or lying just beyond) the BLR. In this case, by assuming \rufo=\routminA\ in Eq. \ref{eq:ekin} \ekin\ would decrease by a factor of $\sim$100. Similarly, by assuming the change in ionization as the cause for B+C component variability over \trest=0.20 yr, we obtain \routminBC $\le$1050 pc. 
Such upper limit is compatible with the typical distance of tens of parsecs found by \cite{He2019} based on the study of absorption troughs variability due to ionization changes of a large sample ($\sim$900) of BAL QSOs. The assumption of \rufo=\routminBC would imply an upper limit estimation on the kinetic power of \ekin$\le$7.9$\times$10$^{45}$ \ergs\ and \ekin$\le$ 5.8$\times$10$^{45}$ \ergs\ for B and C troughs, respectively.

It is therefore evident that to overcome the limitations of present data in accurately constraining  \ekin\ are necessary (i) a spectral monitoring to probe the existence of BAL variability on much shorter timescales than a year and (ii) conclusive information of the presence of PV absorption blueshifted as the CIV trough, which cannot be derived due the low quality of the data in the corresponding spectral region ($\approx$ 930-970 \AA). It is worth noting that a detection of this ion in the UFO would imply a $N\rm_H$ =10$^{22.7}$ \cmdue (e.g. \citealt{Capellupo2014}) and \ekin\ would increase by a factor of 3 dex, high enough to consider these UFOs being playing a key role for an efficient AGN feedback mechanism (i.e. \ekin$\sim$0.01-0.05 $\times$\lbol; \citealt{Faucher2012}, \citealt{DiMatteo2005}.)

We plan to perform a high-resolution (R=40,000) spectral monitoring using instruments such as VLT/UVES and Subaru/HDS on a timescale from days to weeks up to few months, to better constrain physical properties as $N\rm_{H}$, e.g. from PV variability, and \rufo, which will be used for estimating \ekin\ of the UFO in J1538$+$0855. Interestingly, the detection of a BAL variation occurring on \trest\ of $\sim$20 days would imply a distance of \routA$\sim$0.04 pc, i.e. a factor of 10 smaller than the CIV BLR radius in J1538+0855, and \routBC$\sim$550 pc for troughs B and C.
High-resolution spectroscopy is also needed to fully resolve the BAL troughs and enable the identification of blueshifted narrow absorption lines in the UV spectrum, which can trace another phase of the outflows in J1538+0855.
Finally, the presence of such CIV UFO may be considered as a promising mechanism for the origin of the extended ($\sim$75 kpc) CIV nebula surrounding  J1538+0855 discovered by \citealt{Travascio2020}. AGN-driven outflows are indeed invoked to deposit  metal-enriched  gas up large distances from the host galaxy and explain the presence of a non-pristine  circumgalactic medium (e.g., \citealt{Choi2020}). Deep MUSE data will be asked for probing the existence of a  blueshifted component in the emission profile of the CIV nebula  (i.e. similarly to the one discovered for the Ly$\alpha$ nebula) that would support the scenario of QSO-driven CIV outflows that can propagate up to the scale of the CGM.

\begin{acknowledgements}
We are grateful to the anonymous referee for useful feedback which helped us to improve the paper. GV, EP, FF, CF and MB acknowledge support from PRIN MIUR project "Black Hole winds and the Baryon Life Cycle of Galaxies: the stone-guest at the galaxy evolution supper”. EP, LZ and GV acknowledge financial support under ASI-INAF contract 2017-14-H.0. TM acknowledges support from JSPS KAKENHI (Grant Number 21H01126). FN and AL acknowledge support from the European Union’s Horizon 2020 Programme under the AHEAD2020 project (grant agreement n. 871158). The LBT is an international collaboration among institutions in the United States, Italy and Germany. LBT Corporation partners are: Istituto Nazionale di Astrofisica, Italy; The University of Arizona on behalf of the Arizona Board of Regents;  LBT Beteiligungsgesellschaft, Germany, representing the Max-Planck Society, The Leibniz Institute for Astrophysics Potsdam, and Heidelberg University; The Ohio State University, representing OSU, University of Notre Dame, University of Minnesota and University of Virginia. This paper used data obtained with the MODS spectrographs built with funding from NSF grant AST-9987045 and the NSF Telescope System Instrumentation Program (TSIP), with additional funds from the Ohio Board of Regents and the Ohio State University Office of Research. We acknowledge the help of the LBT Italia partnership in accepting our DDT proposal and supporting us throughout the whole program from observations preparation to data reduction. This research is based in part on data collected at the Subaru Telescope, which is operated by the National Astronomical Observatory of Japan. We are honored and grateful for the opportunity of observing the Universe from Maunakea, which has the cultural, historical, and natural significance in Hawaii. Based on data obtained with the European Southern Observatory Very Large Telescope, Paranal, Chile, under Programme 099.A-0316(A). Funding for SDSS-III has been provided by the Alfred P. Sloan Foundation, the Participating Institutions, the National Science Foundation, and the U.S. Department of Energy Office of Science. The SDSS-III web site is http://www.sdss3.org/.
SDSS-III is managed by the Astrophysical Research Consortium for the Participating Institutions of the SDSS-III Collaboration including the University of Arizona, the Brazilian Participation Group, Brookhaven National Laboratory, Carnegie Mellon University, University of Florida, the French Participation Group, the German Participation Group, Harvard University, the Instituto de Astrofisica de Canarias, the Michigan State/Notre Dame/JINA Participation Group, Johns Hopkins University, Lawrence Berkeley National Laboratory, Max Planck Institute for Astrophysics, Max Planck Institute for Extraterrestrial Physics, New Mexico State University, New York University, Ohio State University, Pennsylvania State University, University of Portsmouth, Princeton University, the Spanish Participation Group, University of Tokyo, University of Utah, Vanderbilt University, University of Virginia, University of Washington, and Yale University. 
\end{acknowledgements}

%
%

\bibliographystyle{aa} 
\bibliography{bib} 

\begin{thebibliography}{56}
\expandafter\ifx\csname natexlab\endcsname\relax\def\natexlab#1{#1}\fi

\bibitem[{{Arav} {et~al.}(2018){Arav}, {Liu}, {Xu}, {Stidham}, {Benn}, \&
  {Chamberlain}}]{Arav2018}
{Arav}, N., {Liu}, G., {Xu}, X., {et~al.} 2018, \apj, 857, 60

\bibitem[{{Arnaud} \& {Rothenflug}(1985)}]{Arnaud1985}
{Arnaud}, M. \& {Rothenflug}, R. 1985, \aaps, 60, 425

\bibitem[{{Bischetti} {et~al.}(2019){Bischetti}, {Maiolino}, {Carniani},
  {Fiore}, {Piconcelli}, \& {Fluetsch}}]{Bischetti2019}
{Bischetti}, M., {Maiolino}, R., {Carniani}, S., {et~al.} 2019, \aap, 630, A59

\bibitem[{{Bruni} {et~al.}(2019){Bruni}, {Piconcelli}, {Misawa}, {Zappacosta},
  {Saturni}, {Vietri}, {Vignali}, {Bongiorno}, {Duras}, {Feruglio}, {Tombesi},
  \& {Fiore}}]{Bruni2019}
{Bruni}, G., {Piconcelli}, E., {Misawa}, T., {et~al.} 2019, \aap, 630, A111

\bibitem[{{Capellupo} {et~al.}(2014){Capellupo}, {Hamann}, \&
  {Barlow}}]{Capellupo2014}
{Capellupo}, D.~M., {Hamann}, F., \& {Barlow}, T.~A. 2014, \mnras, 444, 1893

\bibitem[{{Capellupo} {et~al.}(2013){Capellupo}, {Hamann}, {Shields},
  {Halpern}, \& {Barlow}}]{Capellupo2013}
{Capellupo}, D.~M., {Hamann}, F., {Shields}, J.~C., {Halpern}, J.~P., \&
  {Barlow}, T.~A. 2013, \mnras, 429, 1872

\bibitem[{{Capellupo} {et~al.}(2012){Capellupo}, {Hamann}, {Shields},
  {Rodr{\'\i}guez Hidalgo}, \& {Barlow}}]{Capellupo2012}
{Capellupo}, D.~M., {Hamann}, F., {Shields}, J.~C., {Rodr{\'\i}guez Hidalgo},
  P., \& {Barlow}, T.~A. 2012, \mnras, 422, 3249

\bibitem[{{Choi} {et~al.}(2020){Choi}, {Brennan}, {Somerville}, {Ostriker},
  {Hirschmann}, \& {Naab}}]{Choi2020}
{Choi}, E., {Brennan}, R., {Somerville}, R.~S., {et~al.} 2020, \apj, 904, 8

\bibitem[{{Choi} {et~al.}(2018){Choi}, {Somerville}, {Ostriker}, {Naab}, \&
  {Hirschmann}}]{Choi2018}
{Choi}, E., {Somerville}, R.~S., {Ostriker}, J.~P., {Naab}, T., \&
  {Hirschmann}, M. 2018, \apj, 866, 91

\bibitem[{{Cicone} {et~al.}(2018){Cicone}, {Brusa}, {Ramos Almeida}, {Cresci},
  {Husemann}, \& {Mainieri}}]{Cicone2018}
{Cicone}, C., {Brusa}, M., {Ramos Almeida}, C., {et~al.} 2018, Nature
  Astronomy, 2, 176

\bibitem[{{Crenshaw} {et~al.}(2003){Crenshaw}, {Kraemer}, \&
  {George}}]{Crenshaw2003}
{Crenshaw}, D.~M., {Kraemer}, S.~B., \& {George}, I.~M. 2003, \araa, 41, 117

\bibitem[{{De Cicco} {et~al.}(2018){De Cicco}, {Brandt}, {Grier}, {Paolillo},
  {Filiz Ak}, {Schneider}, \& {Trump}}]{Decicco2018}
{De Cicco}, D., {Brandt}, W.~N., {Grier}, C.~J., {et~al.} 2018, \aap, 616, A114

\bibitem[{{Di Matteo} {et~al.}(2005){Di Matteo}, {Springel}, \&
  {Hernquist}}]{DiMatteo2005}
{Di Matteo}, T., {Springel}, V., \& {Hernquist}, L. 2005, \nat, 433, 604

\bibitem[{{Elvis}(2000)}]{Elvis2000}
{Elvis}, M. 2000, \apj, 545, 63

\bibitem[{{Fabian}(2012)}]{Fabian2012}
{Fabian}, A.~C. 2012, \araa, 50, 455

\bibitem[{{Faucher-Gigu{\`e}re} \& {Quataert}(2012)}]{Faucher2012}
{Faucher-Gigu{\`e}re}, C.-A. \& {Quataert}, E. 2012, \mnras, 425, 605

\bibitem[{{Ferland} {et~al.}(2017){Ferland}, {Chatzikos}, {Guzm{\'a}n},
  {Lykins}, {van Hoof}, {Williams}, {Abel}, {Badnell}, {Keenan}, {Porter}, \&
  {Stancil}}]{Ferland2017}
{Ferland}, G.~J., {Chatzikos}, M., {Guzm{\'a}n}, F., {et~al.} 2017, \rmxaa, 53,
  385

\bibitem[{{Ferland} {et~al.}(1998){Ferland}, {Korista}, {Verner}, {Ferguson},
  {Kingdon}, \& {Verner}}]{Ferland1998}
{Ferland}, G.~J., {Korista}, K.~T., {Verner}, D.~A., {et~al.} 1998, \pasp, 110,
  761

\bibitem[{{Filiz Ak} {et~al.}(2012){Filiz Ak}, {Brandt}, {Hall}, {Schneider},
  {Anderson}, {Gibson}, {Lundgren}, {Myers}, {Petitjean}, {Ross}, {Shen},
  {York}, {Bizyaev}, {Brinkmann}, {Malanushenko}, {Oravetz}, {Pan}, {Simmons},
  \& {Weaver}}]{Filizak2012}
{Filiz Ak}, N., {Brandt}, W.~N., {Hall}, P.~B., {et~al.} 2012, \apj, 757, 114

\bibitem[{{Filiz Ak} {et~al.}(2013){Filiz Ak}, {Brandt}, {Hall}, {Schneider},
  {Anderson}, {Hamann}, {Lundgren}, {Myers}, {P{\^a}ris}, {Petitjean}, {Ross},
  {Shen}, \& {York}}]{FilizAk2013}
{Filiz Ak}, N., {Brandt}, W.~N., {Hall}, P.~B., {et~al.} 2013, \apj, 777, 168

\bibitem[{{Fiore} {et~al.}(2017){Fiore}, {Feruglio}, {Shankar}, {Bischetti},
  {Bongiorno}, {Brusa}, {Carniani}, {Cicone}, {Duras}, {Lamastra}, {Mainieri},
  {Marconi}, {Menci}, {Maiolino}, {Piconcelli}, {Vietri}, \&
  {Zappacosta}}]{Fiore2017}
{Fiore}, F., {Feruglio}, C., {Shankar}, F., {et~al.} 2017, \aap, 601, A143

\bibitem[{{Fukumura} {et~al.}(2015){Fukumura}, {Tombesi}, {Kazanas}, {Shrader},
  {Behar}, \& {Contopoulos}}]{Fukumura2015}
{Fukumura}, K., {Tombesi}, F., {Kazanas}, D., {et~al.} 2015, \apj, 805, 17

\bibitem[{{Ganguly} {et~al.}(2001){Ganguly}, {Bond}, {Charlton}, {Eracleous},
  {Brandt}, \& {Churchill}}]{Ganguly2001}
{Ganguly}, R., {Bond}, N.~A., {Charlton}, J.~C., {et~al.} 2001, \apj, 549, 133

\bibitem[{{Gibson} {et~al.}(2008){Gibson}, {Brandt}, {Schneider}, \&
  {Gallagher}}]{Gibson2008}
{Gibson}, R.~R., {Brandt}, W.~N., {Schneider}, D.~P., \& {Gallagher}, S.~C.
  2008, \apj, 675, 985

\bibitem[{{Gibson} {et~al.}(2009){Gibson}, {Jiang}, {Brandt}, {Hall}, {Shen},
  {Wu}, {Anderson}, {Schneider}, {Vanden Berk}, {Gallagher}, {Fan}, \&
  {York}}]{Gibson2009}
{Gibson}, R.~R., {Jiang}, L., {Brandt}, W.~N., {et~al.} 2009, \apj, 692, 758

\bibitem[{{Gofford} {et~al.}(2015){Gofford}, {Reeves}, {McLaughlin}, {Braito},
  {Turner}, {Tombesi}, \& {Cappi}}]{Gofford2015}
{Gofford}, J., {Reeves}, J.~N., {McLaughlin}, D.~E., {et~al.} 2015, \mnras,
  451, 4169

\bibitem[{{Gofford} {et~al.}(2013){Gofford}, {Reeves}, {Tombesi}, {Braito},
  {Turner}, {Miller}, \& {Cappi}}]{Gofford2013}
{Gofford}, J., {Reeves}, J.~N., {Tombesi}, F., {et~al.} 2013, \mnras, 430, 60

\bibitem[{{Gunn} {et~al.}(2006){Gunn}, {Siegmund}, {Mannery}, {Owen}, {Hull},
  {Leger}, {Carey}, {Knapp}, {York}, {Boroski}, {Kent}, {Lupton}, {Rockosi},
  {Evans}, {Waddell}, {Anderson}, {Annis}, {Barentine}, {Bartoszek}, {Bastian},
  {Bracker}, {Brewington}, {Briegel}, {Brinkmann}, {Brown}, {Carr},
  {Czarapata}, {Drennan}, {Dombeck}, {Federwitz}, {Gillespie}, {Gonzales},
  {Hansen}, {Harvanek}, {Hayes}, {Jordan}, {Kinney}, {Klaene}, {Kleinman},
  {Kron}, {Kresinski}, {Lee}, {Limmongkol}, {Lindenmeyer}, {Long}, {Loomis},
  {McGehee}, {Mantsch}, {Neilsen}, {Neswold}, {Newman}, {Nitta}, {Peoples},
  {Pier}, {Prieto}, {Prosapio}, {Rivetta}, {Schneider}, {Snedden}, \&
  {Wang}}]{Gunn2006}
{Gunn}, J.~E., {Siegmund}, W.~A., {Mannery}, E.~J., {et~al.} 2006, \aj, 131,
  2332

\bibitem[{{Hamann} {et~al.}(2018){Hamann}, {Chartas}, {Reeves}, \&
  {Nardini}}]{Hamann2018}
{Hamann}, F., {Chartas}, G., {Reeves}, J., \& {Nardini}, E. 2018, \mnras, 476,
  943

\bibitem[{{Hamann} {et~al.}(2019){Hamann}, {Herbst}, {Paris}, \&
  {Capellupo}}]{Hamann2019}
{Hamann}, F., {Herbst}, H., {Paris}, I., \& {Capellupo}, D. 2019, \mnras, 483,
  1808

\bibitem[{{Harrison} {et~al.}(2018){Harrison}, {Costa}, {Tadhunter},
  {Fl{\"u}tsch}, {Kakkad}, {Perna}, \& {Vietri}}]{Harrison2018}
{Harrison}, C.~M., {Costa}, T., {Tadhunter}, C.~N., {et~al.} 2018, Nature
  Astronomy, 2, 198

\bibitem[{{He} {et~al.}(2019){He}, {Wang}, {Liu}, {Wang}, {Bian},
  {Tchernyshyov}, {Mou}, {Xu}, {Zhou}, {Green}, \& {Xu}}]{He2019}
{He}, Z., {Wang}, T., {Liu}, G., {et~al.} 2019, Nature Astronomy, 3, 265

\bibitem[{{Lira} {et~al.}(2018){Lira}, {Kaspi}, {Netzer}, {Botti}, {Morrell},
  {Mej{\'\i}a-Restrepo}, {S{\'a}nchez-S{\'a}ez}, {Mart{\'\i}nez-Palomera}, \&
  {L{\'o}pez}}]{Lira2018}
{Lira}, P., {Kaspi}, S., {Netzer}, H., {et~al.} 2018, \apj, 865, 56

\bibitem[{{Lodders}(2003)}]{Lodders2003}
{Lodders}, K. 2003, \apj, 591, 1220

\bibitem[{{Luminari} {et~al.}(2020){Luminari}, {Tombesi}, {Piconcelli},
  {Nicastro}, {Fukumura}, {Kazanas}, {Fiore}, \& {Zappacosta}}]{Luminari2020}
{Luminari}, A., {Tombesi}, F., {Piconcelli}, E., {et~al.} 2020, \aap, 633, A55

\bibitem[{{Miller} {et~al.}(2020){Miller}, {Arav}, {Xu}, \&
  {Kriss}}]{Miller2020}
{Miller}, T.~R., {Arav}, N., {Xu}, X., \& {Kriss}, G.~A. 2020, \mnras, 499,
  1522

\bibitem[{{Morgan} {et~al.}(2010){Morgan}, {Kochanek}, {Morgan}, \&
  {Falco}}]{Morgan2010}
{Morgan}, C.~W., {Kochanek}, C.~S., {Morgan}, N.~D., \& {Falco}, E.~E. 2010,
  \apj, 712, 1129

\bibitem[{{Murray} {et~al.}(1995){Murray}, {Chiang}, {Grossman}, \&
  {Voit}}]{Murray1995}
{Murray}, N., {Chiang}, J., {Grossman}, S.~A., \& {Voit}, G.~M. 1995, \apj,
  451, 498

\bibitem[{{Nagao} {et~al.}(2006){Nagao}, {Marconi}, \& {Maiolino}}]{Nagao2006}
{Nagao}, T., {Marconi}, A., \& {Maiolino}, R. 2006, \aap, 447, 157

\bibitem[{{Narayanan} {et~al.}(2004){Narayanan}, {Hamann}, {Barlow},
  {Burbidge}, {Cohen}, {Junkkarinen}, \& {Lyons}}]{Narayanan2004}
{Narayanan}, D., {Hamann}, F., {Barlow}, T., {et~al.} 2004, \apj, 601, 715

\bibitem[{{Peterson}(1997)}]{Peterson1997}
{Peterson}, B.~M. 1997, {An Introduction to Active Galactic Nuclei}

\bibitem[{{Proga}(2007)}]{Proga2007}
{Proga}, D. 2007, \apj, 661, 693

\bibitem[{{Proga} {et~al.}(2000){Proga}, {Stone}, \& {Kallman}}]{Proga2000}
{Proga}, D., {Stone}, J.~M., \& {Kallman}, T.~R. 2000, \apj, 543, 686

\bibitem[{{Rodr{\'\i}guez Hidalgo} {et~al.}(2020){Rodr{\'\i}guez Hidalgo},
  {Khatri}, {Hall}, {Haas}, {Quintero}, {Khatu}, {Kowash}, \&
  {Murray}}]{Rodriguez2020}
{Rodr{\'\i}guez Hidalgo}, P., {Khatri}, A.~M., {Hall}, P.~B., {et~al.} 2020,
  \apj, 896, 151

\bibitem[{{Savage} \& {Sembach}(1991)}]{Savage1991}
{Savage}, B.~D. \& {Sembach}, K.~R. 1991, \apj, 379, 245

\bibitem[{{Silk} \& {Rees}(1998)}]{Silk1998}
{Silk}, J. \& {Rees}, M.~J. 1998, \aap, 331, L1

\bibitem[{{Smith} {et~al.}(2019){Smith}, {Tombesi}, {Veilleux}, {Lohfink}, \&
  {Luminari}}]{Smith2019}
{Smith}, R.~N., {Tombesi}, F., {Veilleux}, S., {Lohfink}, A.~M., \& {Luminari},
  A. 2019, \apj, 887, 69

\bibitem[{{Tombesi} {et~al.}(2012){Tombesi}, {Cappi}, {Reeves}, \&
  {Braito}}]{Tombesi2012}
{Tombesi}, F., {Cappi}, M., {Reeves}, J.~N., \& {Braito}, V. 2012, \mnras, 422,
  L1

\bibitem[{{Tombesi} {et~al.}(2010){Tombesi}, {Cappi}, {Reeves}, {Palumbo},
  {Yaqoob}, {Braito}, \& {Dadina}}]{Tombesi2010}
{Tombesi}, F., {Cappi}, M., {Reeves}, J.~N., {et~al.} 2010, \aap, 521, A57

\bibitem[{{Tombesi} {et~al.}(2015){Tombesi}, {Mel{\'e}ndez}, {Veilleux},
  {Reeves}, {Gonz{\'a}lez-Alfonso}, \& {Reynolds}}]{Tombesi2015}
{Tombesi}, F., {Mel{\'e}ndez}, M., {Veilleux}, S., {et~al.} 2015, \nat, 519,
  436

\bibitem[{{Travascio} {et~al.}(2020){Travascio}, {Zappacosta}, {Cantalupo},
  {Piconcelli}, {Arrigoni Battaia}, {Ginolfi}, {Bischetti}, {Vietri},
  {Bongiorno}, {D'Odorico}, {Duras}, {Feruglio}, {Vignali}, \&
  {Fiore}}]{Travascio2020}
{Travascio}, A., {Zappacosta}, L., {Cantalupo}, S., {et~al.} 2020, \aap, 635,
  A157

\bibitem[{{Trevese} {et~al.}(2013){Trevese}, {Saturni}, {Vagnetti}, {Perna},
  {Paris}, \& {Turriziani}}]{Trevese2013}
{Trevese}, D., {Saturni}, F.~G., {Vagnetti}, F., {et~al.} 2013, \aap, 557, A91

\bibitem[{{Veilleux} {et~al.}(2020){Veilleux}, {Maiolino}, {Bolatto}, \&
  {Aalto}}]{Veilleux2020}
{Veilleux}, S., {Maiolino}, R., {Bolatto}, A.~D., \& {Aalto}, S. 2020, \aapr,
  28, 2

\bibitem[{{Vietri} {et~al.}(2018){Vietri}, {Piconcelli}, {Bischetti}, {Duras},
  {Martocchia}, {Bongiorno}, {Marconi}, {Zappacosta}, {Bisogni}, {Bruni},
  {Brusa}, {Comastri}, {Cresci}, {Feruglio}, {Giallongo}, {La Franca},
  {Mainieri}, {Mannucci}, {Ricci}, {Sani}, {Testa}, {Tombesi}, {Vignali}, \&
  {Fiore}}]{Vietri2018}
{Vietri}, G., {Piconcelli}, E., {Bischetti}, M., {et~al.} 2018, \aap, 617, A81

\bibitem[{{Weymann} {et~al.}(1991){Weymann}, {Morris}, {Foltz}, \&
  {Hewett}}]{Weymann1991}
{Weymann}, R.~J., {Morris}, S.~L., {Foltz}, C.~B., \& {Hewett}, P.~C. 1991,
  \apj, 373, 23

\bibitem[{{Zubovas} \& {King}(2012)}]{Zubovas2012}
{Zubovas}, K. \& {King}, A. 2012, \apjl, 745, L34

\end{thebibliography}

\onecolumn

\appendix



\newpage

\section{RH20 continuum normalization}\label{sec:appendix1}
   The method used by RH20 to normalize the quasar spectra is based on a power law anchored to the value of the continuum in three well-defined spectral regions. Specifically, RH20 used the regions of 1701-1725 \AA\ (R$_1$) and 1677-1701 \AA\ (R$_2$), and 1280-1284 \AA\ (R$_3$) or 1415-1430 \AA\ (R$_4$) regions to define the slope. Since both region R$_3$ and R$_4$ can be affected by emission and/or absorption features, they firstly fit a power-law by using R$_1$, R$_2$ and R$_3$ ranges. They compare the result at the midpoint of region R$_4$, taking as good those slopes falling within three times the median error on the flux in region R$_4$, otherwise they anchored the slopes using R$_1$, R$_2$ and R$_4$ regions. Fig. \ref{fig:B1} shows the application of the RH20 continuum normalization method to the spectra of J1538+0855 taken in 2006, 2012, 2017, 2018 and 2021.
   
 \noindent\begin{minipage}{\textwidth}
 \begin{minipage}{.5\textwidth}
   \centering
   \includegraphics[width=8cm]{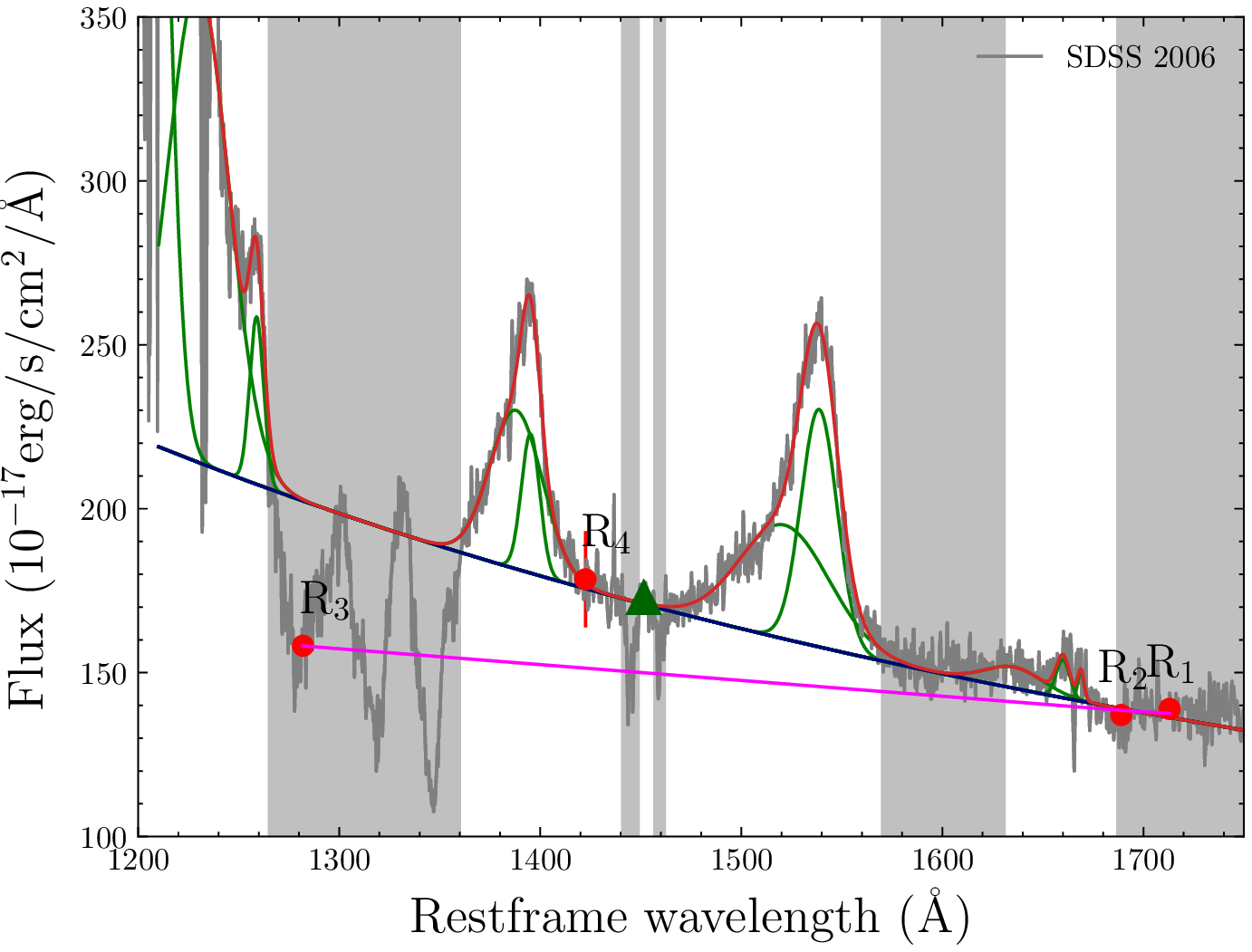}
      \end{minipage}%
 \begin{minipage}{.5\textwidth}
   \centering
   \includegraphics[width=8cm]{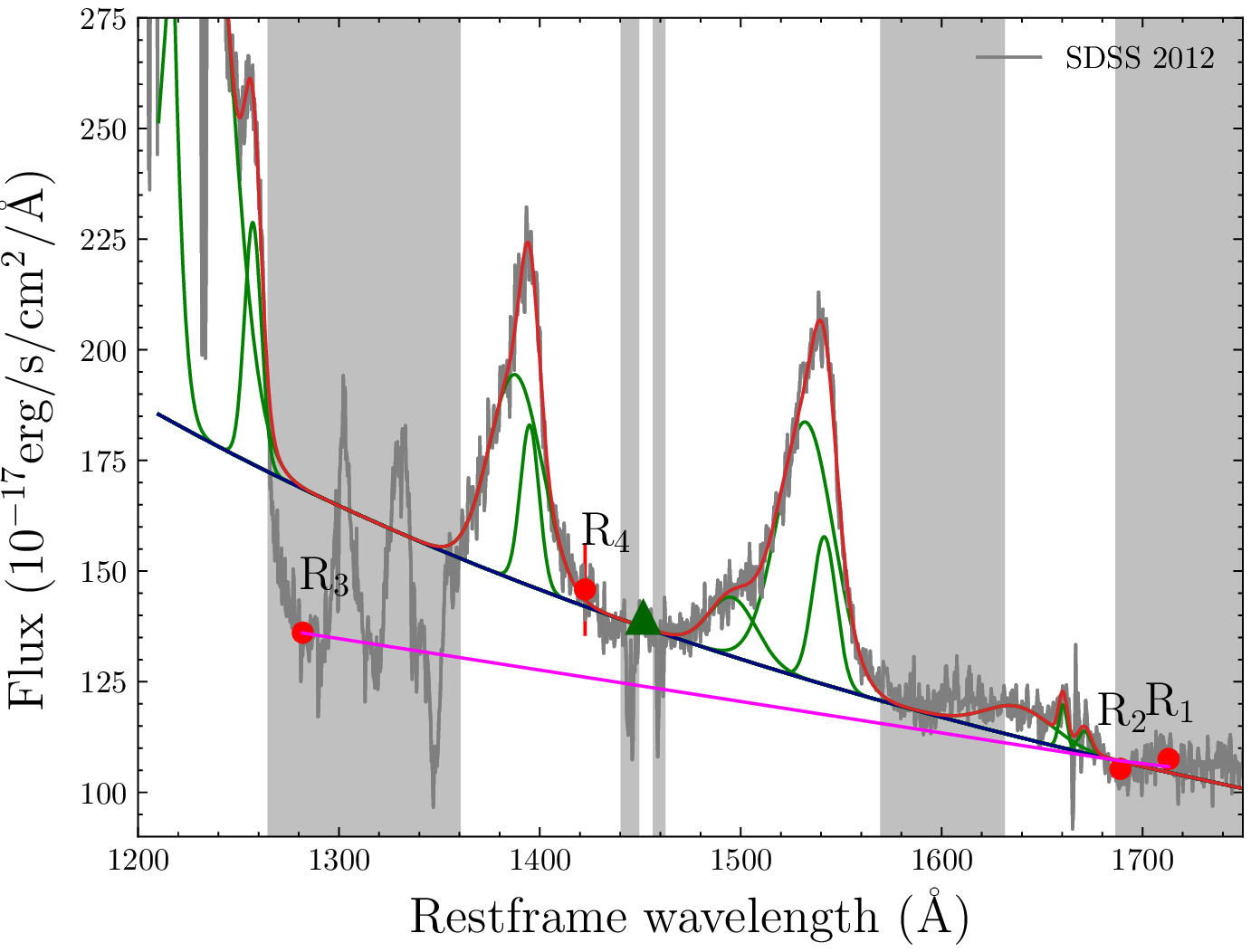}
      \end{minipage}\\
 \begin{minipage}{.5\textwidth}
   \centering
   \includegraphics[width=8cm]{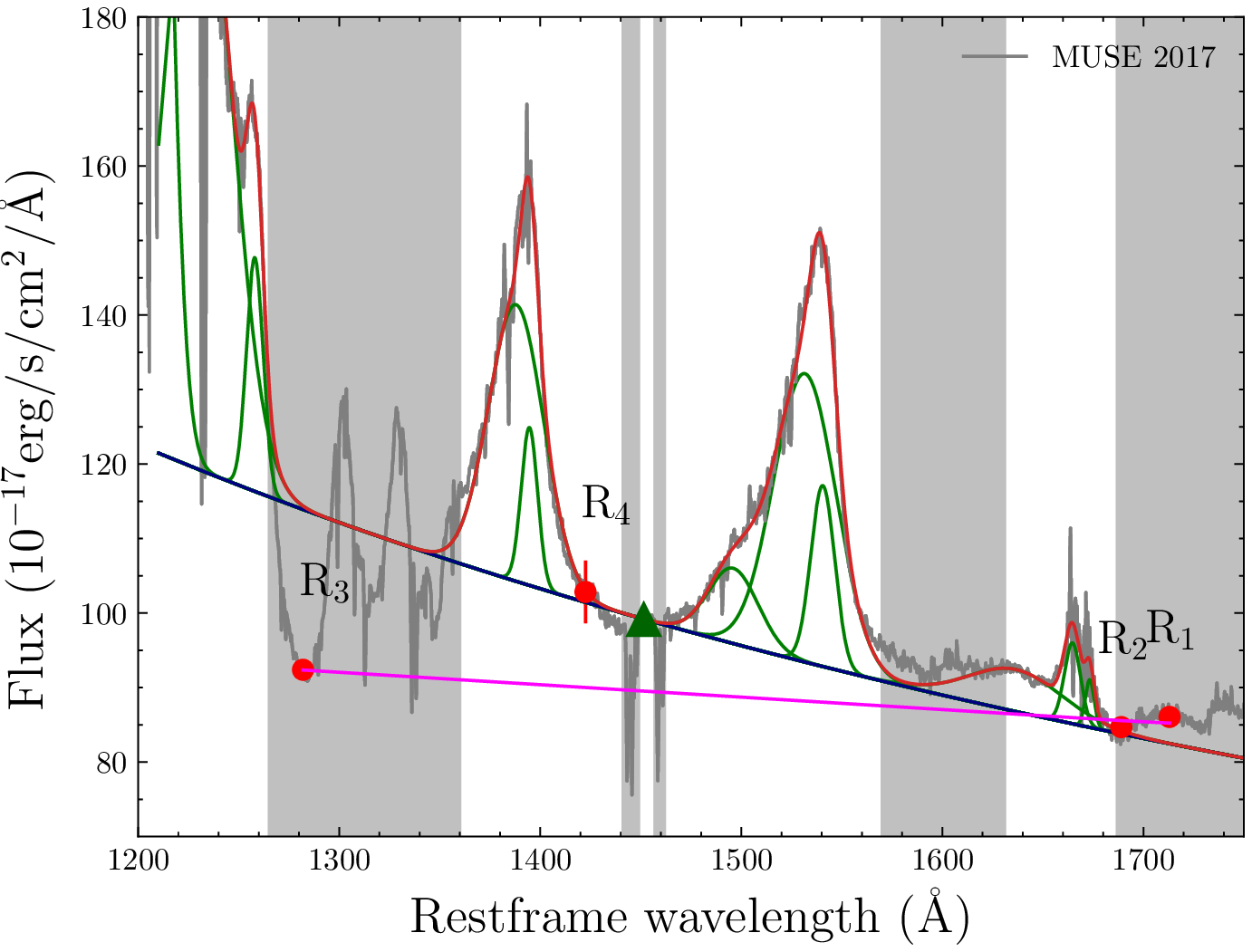}
      \end{minipage}
 \begin{minipage}{.5\textwidth}
   \centering
   \includegraphics[width=8cm]{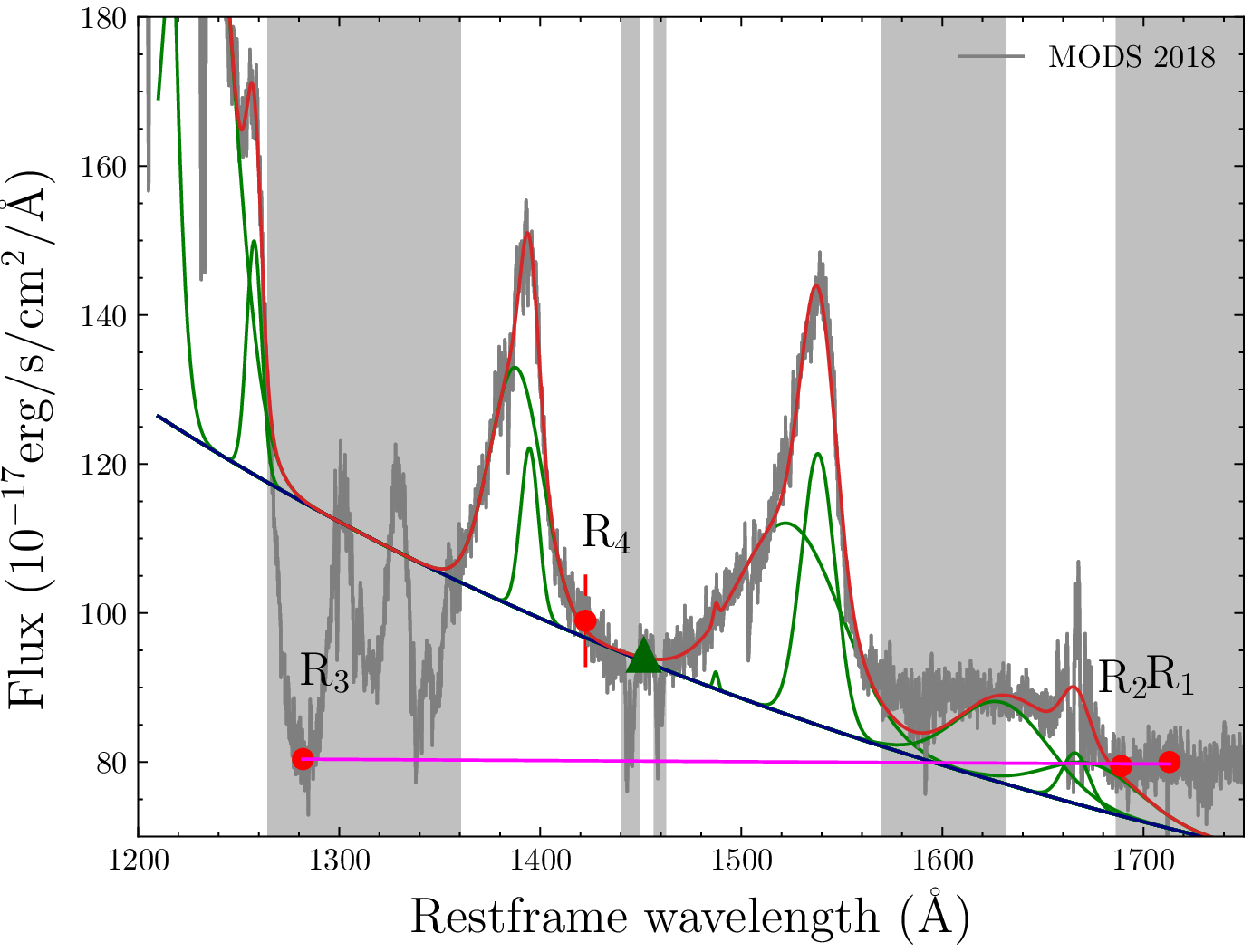}
      \end{minipage}\\
 \begin{minipage}{.5\textwidth}
   \centering
    \includegraphics[width=8cm]{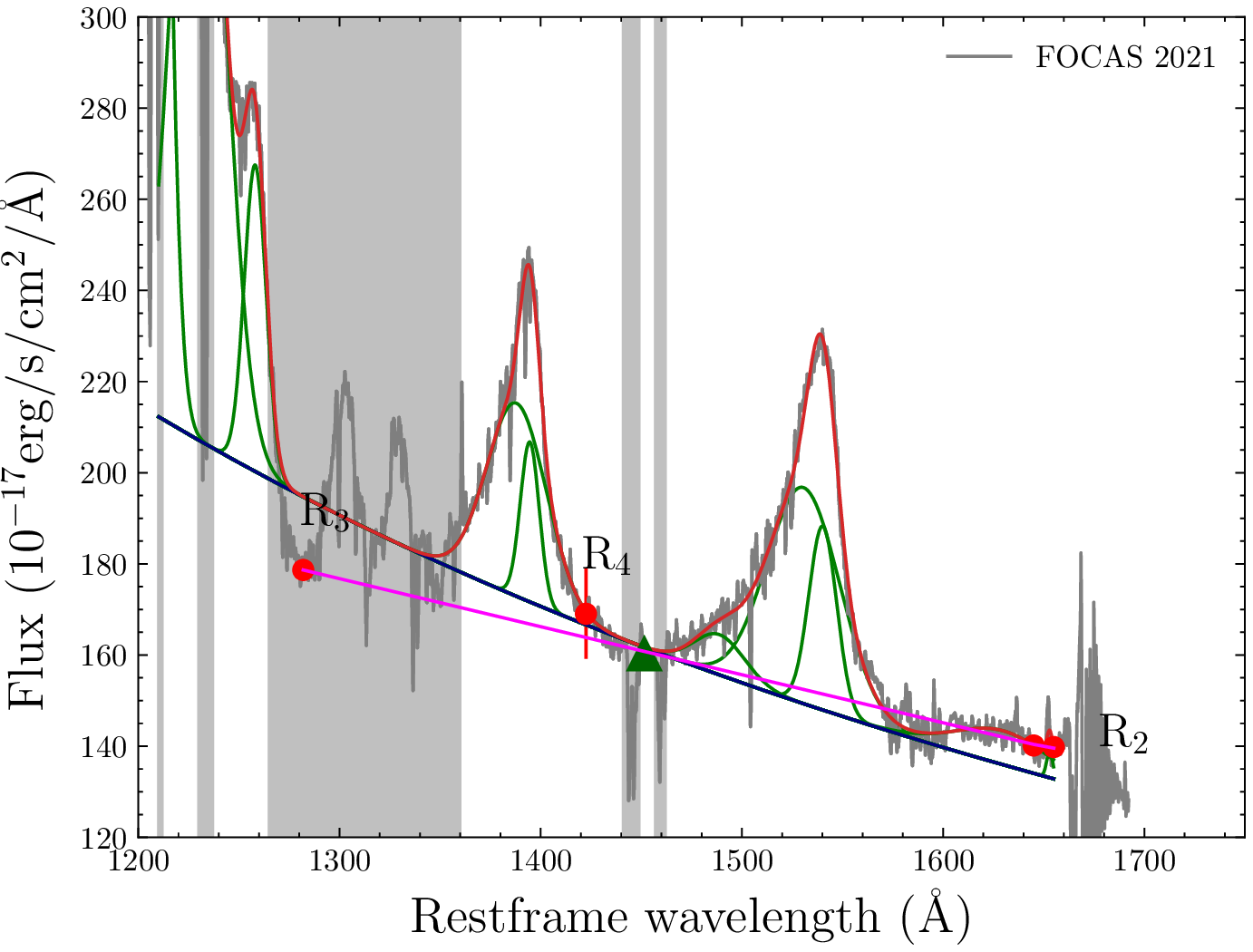}
   \end{minipage}%
   \captionof{figure}{Spectra of J1538+0855 (grey), showing our best-fit model for each epoch (red). The grey bands represent the regions masked during the continuum and emission lines fit. We fit a power law (blue) to the spectrum and Gaussian functions for the emission lines (green curves). The best-fit power-law derived following the method of RH20 is shown as magenta solid line, along with points R$_1$ (median flux of points with rest-frame wavelengths between 1701–1725 \AA), R$_2$ (1677–1701 \AA), R$_3$ (1280–1284 \AA) and R$_4$ (1415–1430 \AA) used to define and anchor the power law slope. The green triangle is used as reference for the continuum level since the R$_4$ point fall within the red-end of the SiIV emission line.}\label{fig:B1}
   \end{minipage}%

  

   

\newpage
\section{Detection of CIV BAL UFOs}\label{sec:appendix2}

In this Appendix, the continuum-normalized spectra of J1538+0855 taken at different epochs (See Table \ref{table:1}) are presented.  Fig. \ref{fig:C1} shows the rest-frame wavelength region bluewards of the SiIV emission line with the CIV BAL UFO troughs at the corresponding velocities from 2012, 2017, 2018 and 2021 spectra.

 \noindent\begin{minipage}{\textwidth}
 \begin{minipage}{.5\textwidth}
   \centering
   \includegraphics[width=8cm]{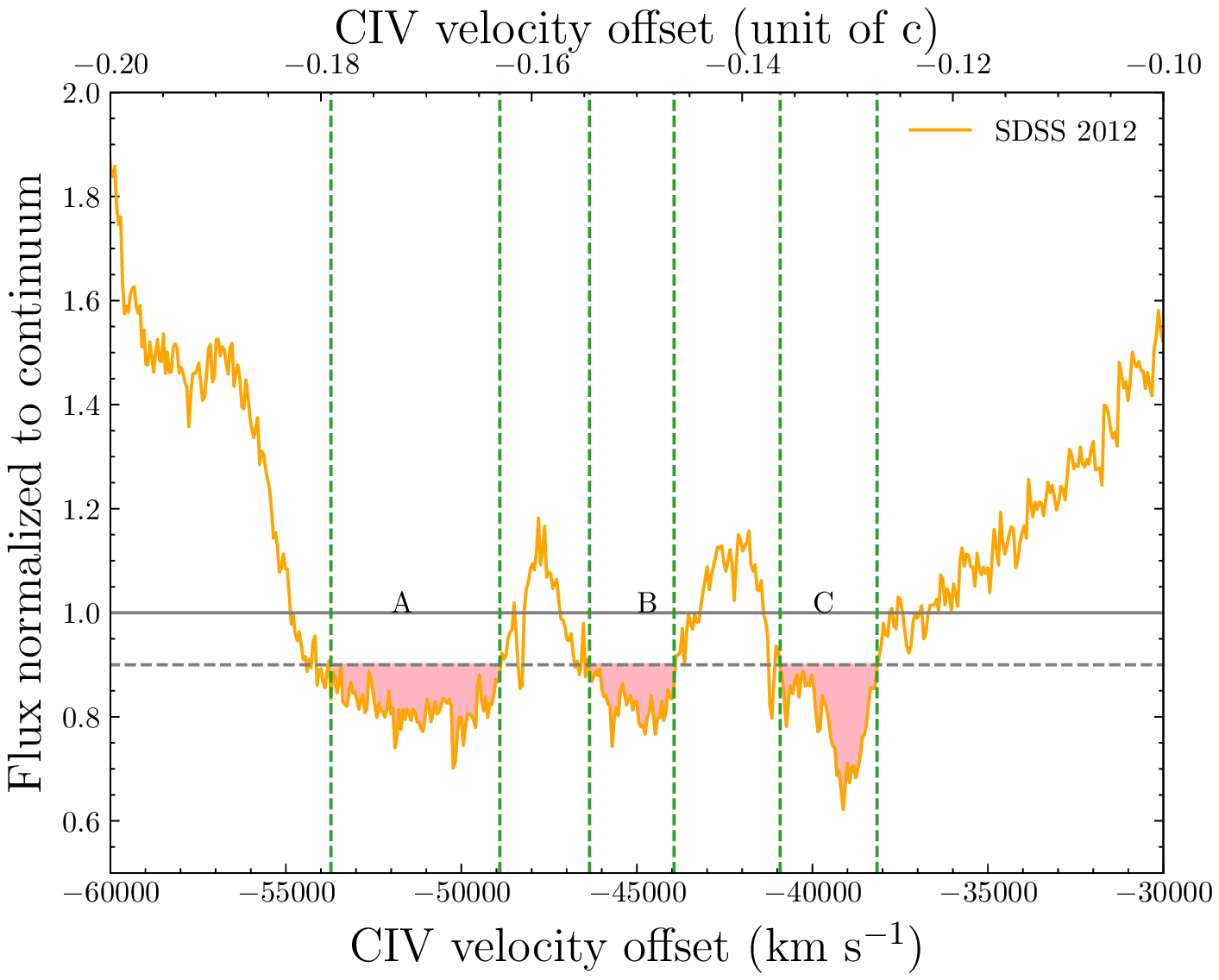}
   \end{minipage}%
 \begin{minipage}{.5\textwidth}
   \centering
  \includegraphics[width=8cm]{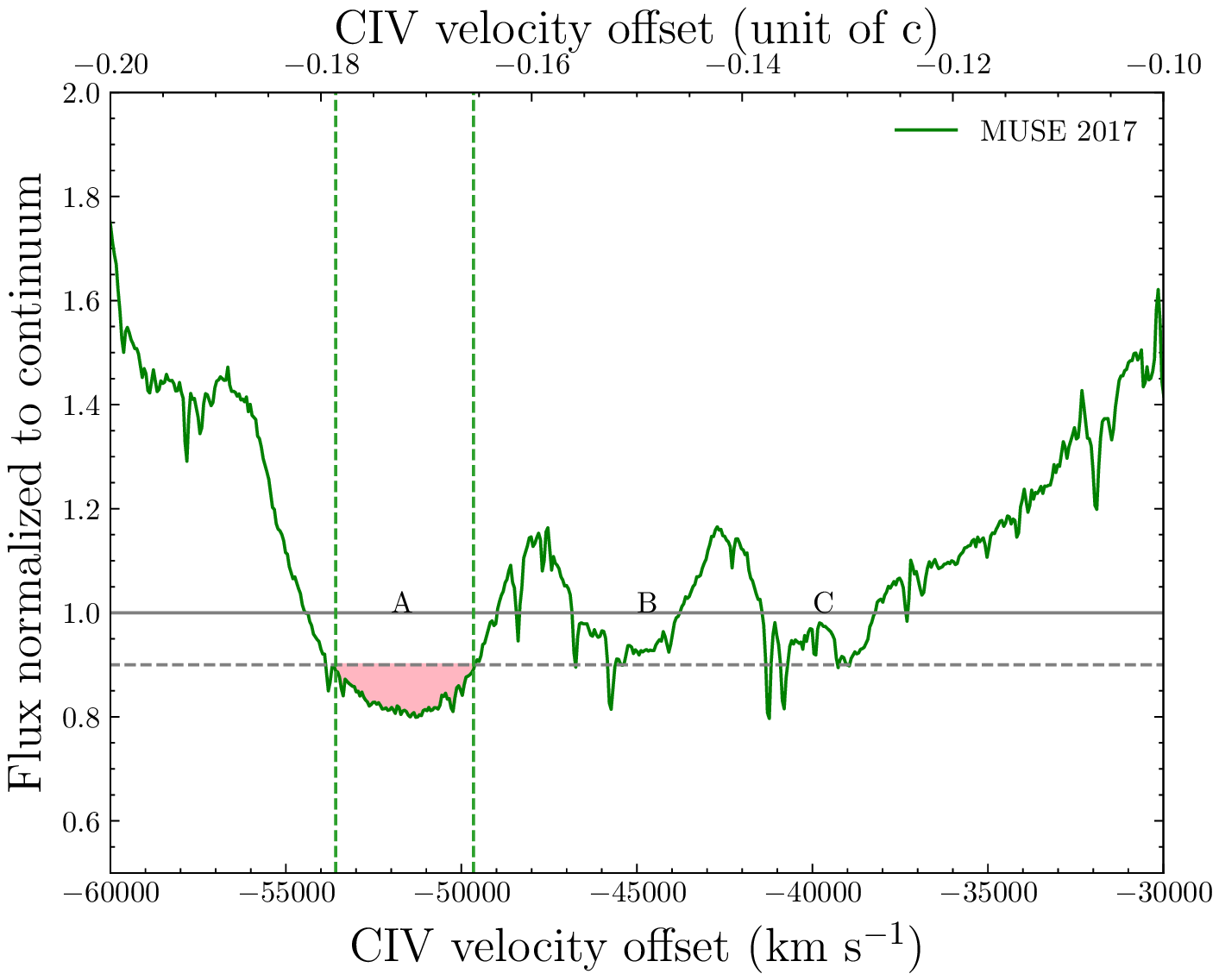}
\end{minipage}\\
 \begin{minipage}{.5\textwidth}
   \centering
   \includegraphics[width=8cm]{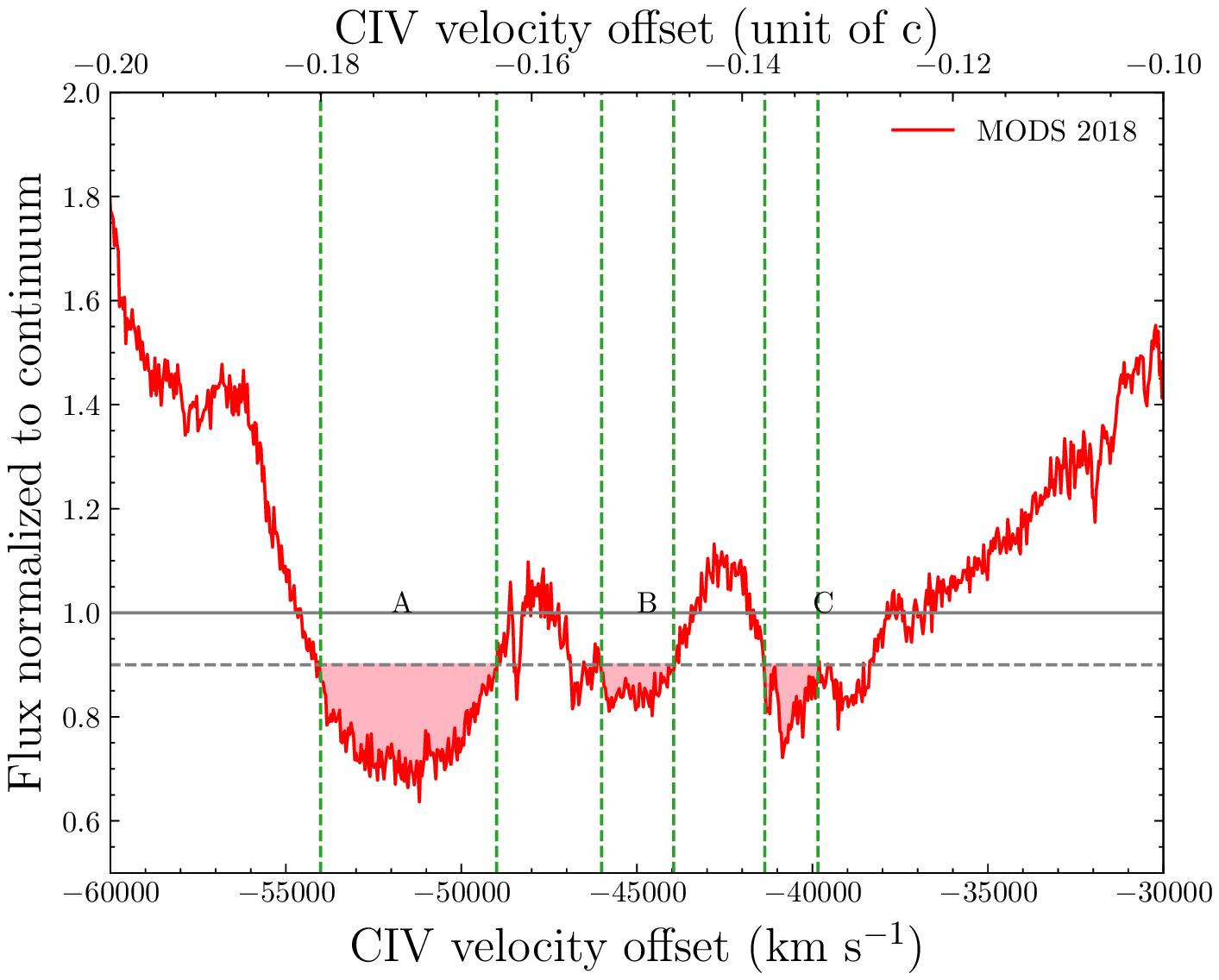}
\end{minipage}%
 \begin{minipage}{.5\textwidth}
   \centering
   \includegraphics[width=8cm]{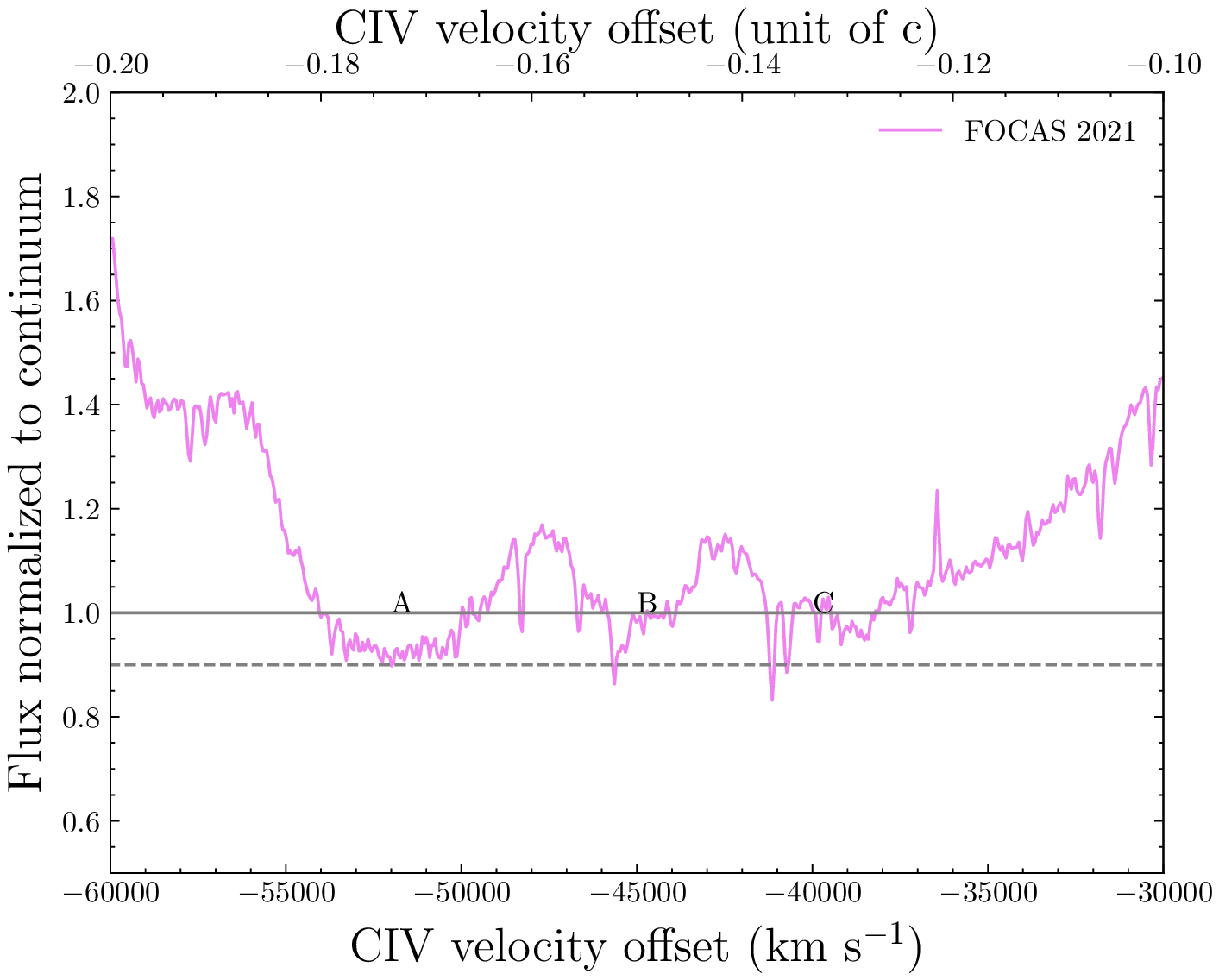}
\end{minipage}%
\captionof{figure}{Light pink regions represent the absorptions below 90\% of the continuum. Green dashed lines indicate the minimum and maximum velocity estimated for the BAL outflow in each trough. Horizontal dashed line represents the 90\% level of the continuum-normalized flux.}\label{fig:C1}
\end{minipage}


\newpage
\section{Definition of BAL variability}\label{sec:appendix3}

In this appendix, comparisons of the continuum-normalized spectra of J1538+0855 between two consecutive epochs are presented. Fig. \ref{fig:D1_A} shows the CIV A trough comparison along with the velocity interval varied between two epochs, marked as grey band, and defined to be at least 1200 km s$^{-1}$ in width with the flux difference in this region be at least 4$\sigma$. We derived for each epoch, within this velocity interval, the \textit{A$_S$} parameter, defined as the fraction of the normalized continuum flux removed by the absorption.  Fig. \ref{fig:D1_B} and \ref{fig:D1_C} show the same as Fig. \ref{fig:D1_A} but for B and C troughs velocity interval, respectively.

 \noindent\begin{minipage}{\textwidth}
  \begin{minipage}{.5\textwidth}
   \centering
    \includegraphics[width=7cm]{Trough_A_variability_DR10_J1538_DR14_J1538_last.eps}
         \end{minipage}%
 \begin{minipage}{.5\textwidth}
   \centering
   \includegraphics[width=7cm]{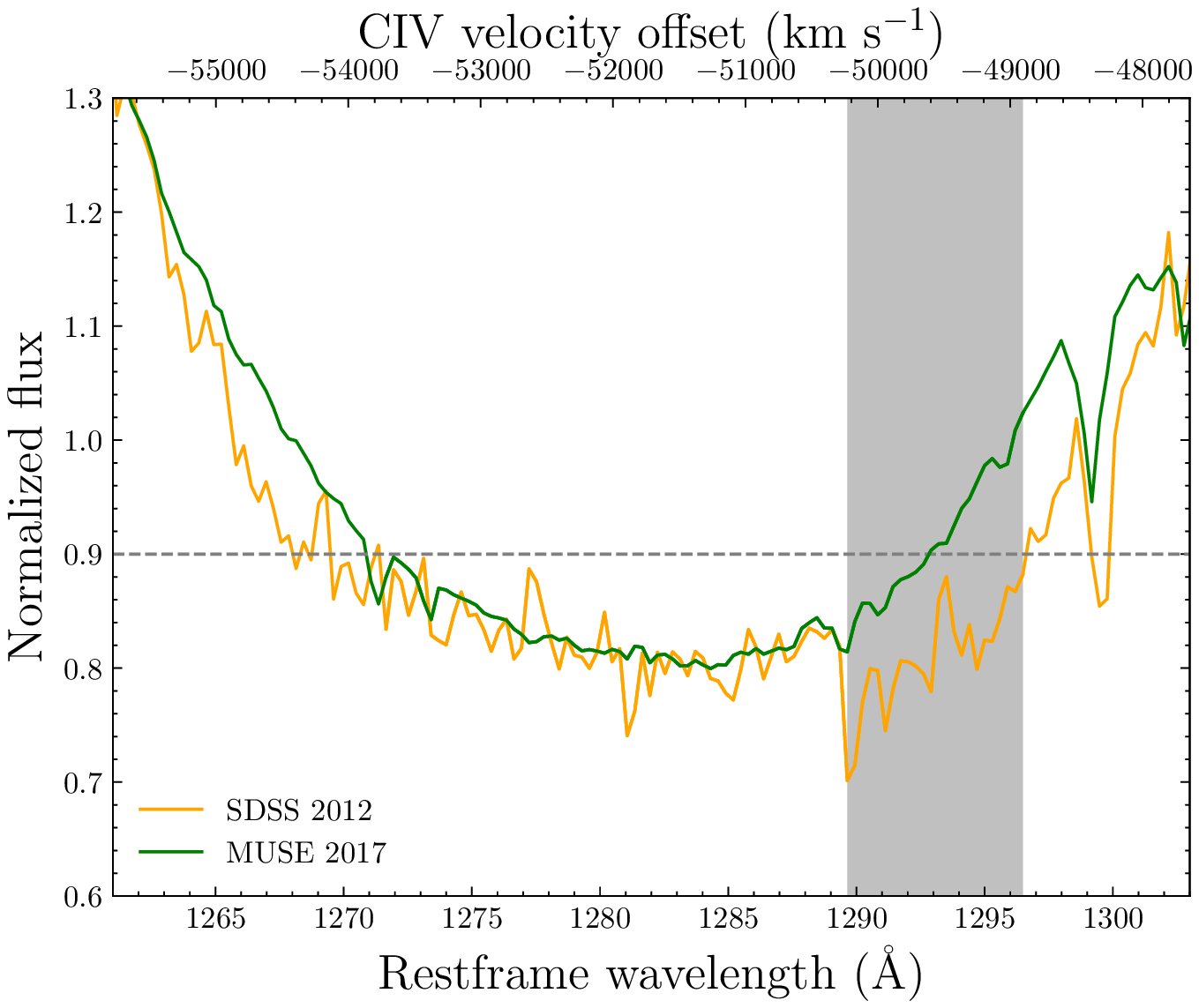}
         \end{minipage}\\
 \begin{minipage}{.5\textwidth}
   \centering
   \includegraphics[width=7cm]{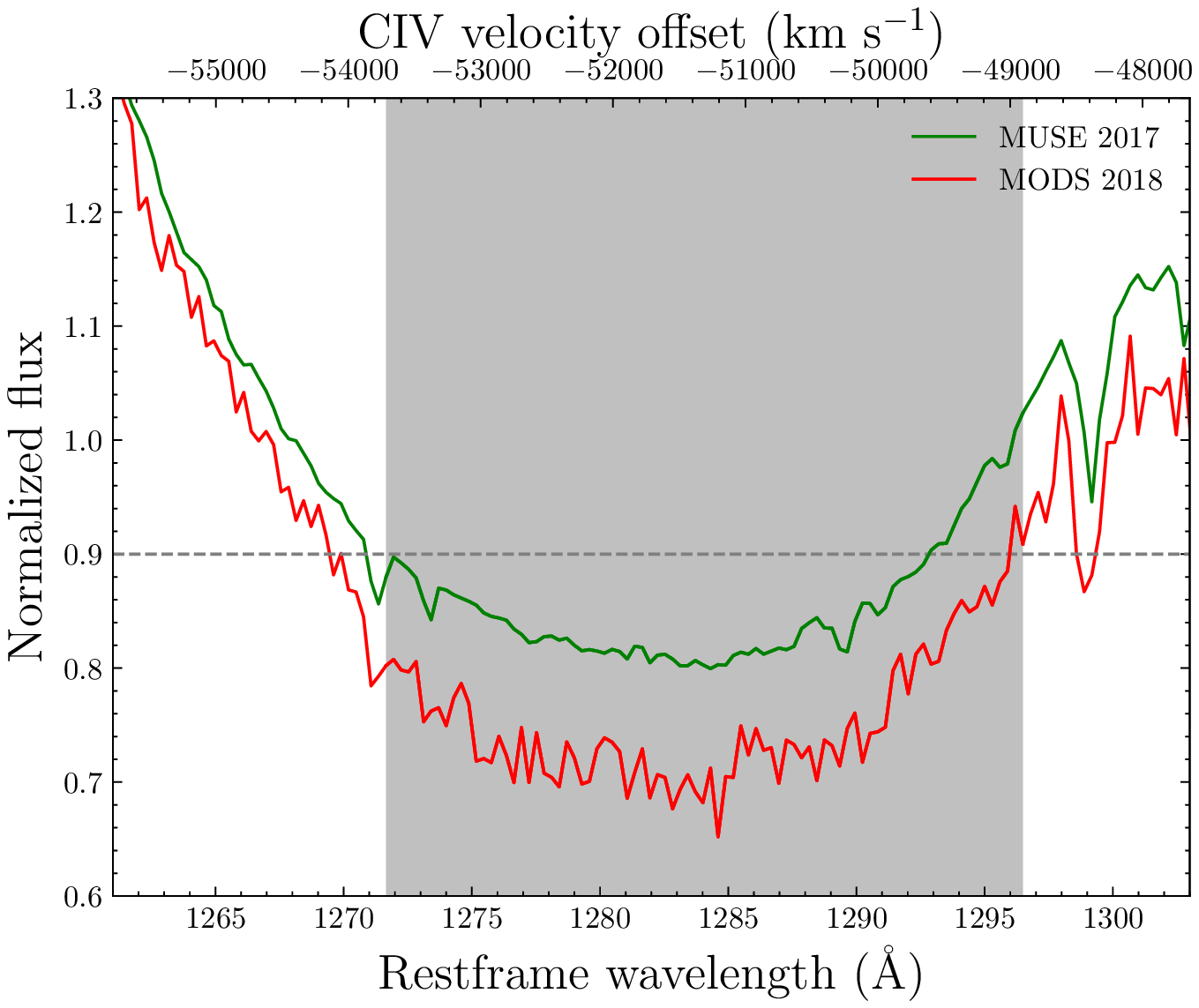}
         \end{minipage}
 \begin{minipage}{.5\textwidth}
   \centering
   \includegraphics[width=7cm]{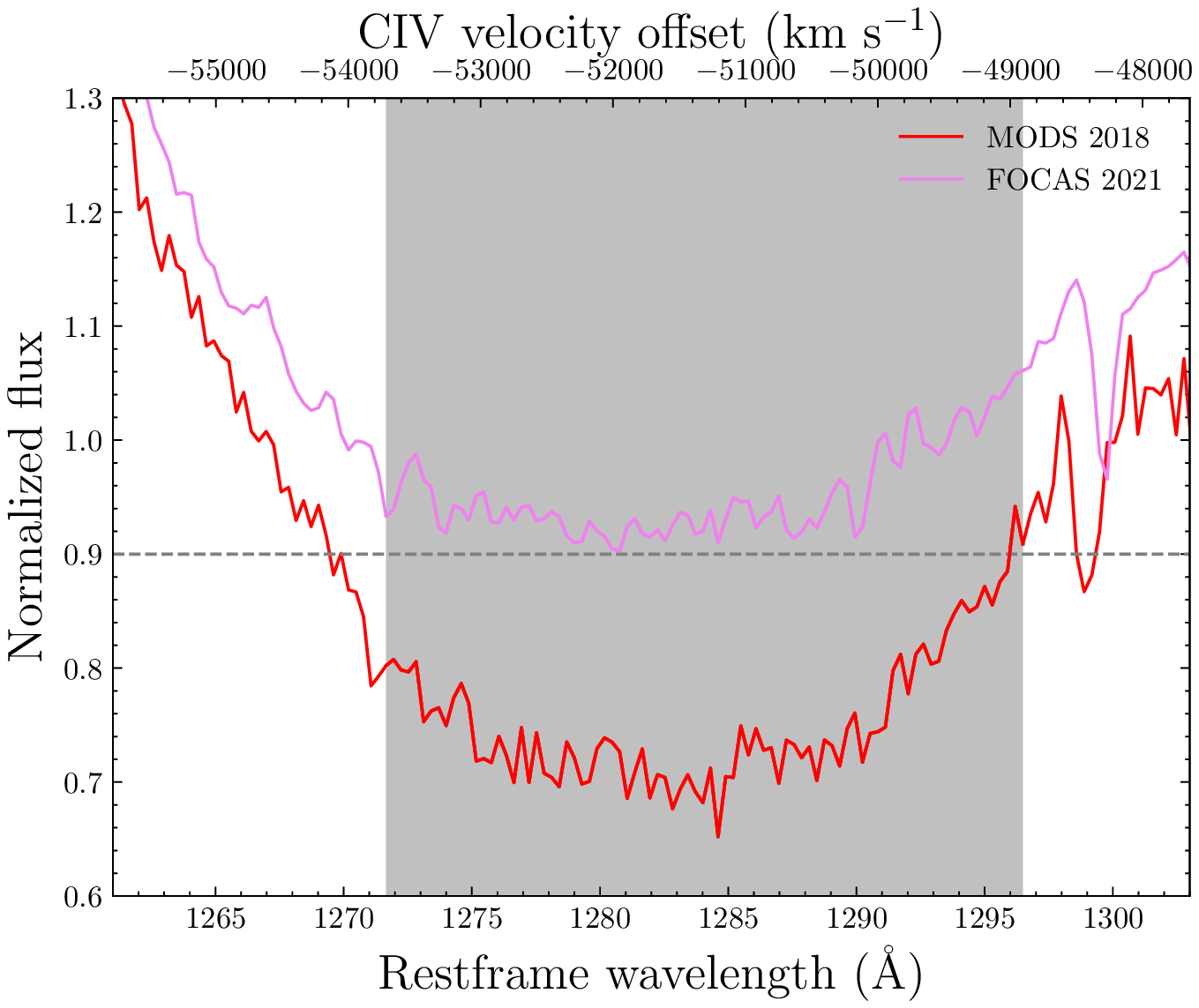}
         \end{minipage}%
\captionof{figure}{Comparison of spectra taken at two consecutive epochs over the velocity interval of the CIV A trough. Shaded region marks where the spectrum varies between the two epochs, which is used to define the region within the \textit{A$_s$} absorption strength is measured. Dashed line represents the 90\% level of the continuum-normalized flux.}\label{fig:D1_A}
\end{minipage}%

\newpage

 \noindent\begin{minipage}{\textwidth}
 \begin{minipage}{.5\textwidth}
   \centering
   \includegraphics[width=7cm]{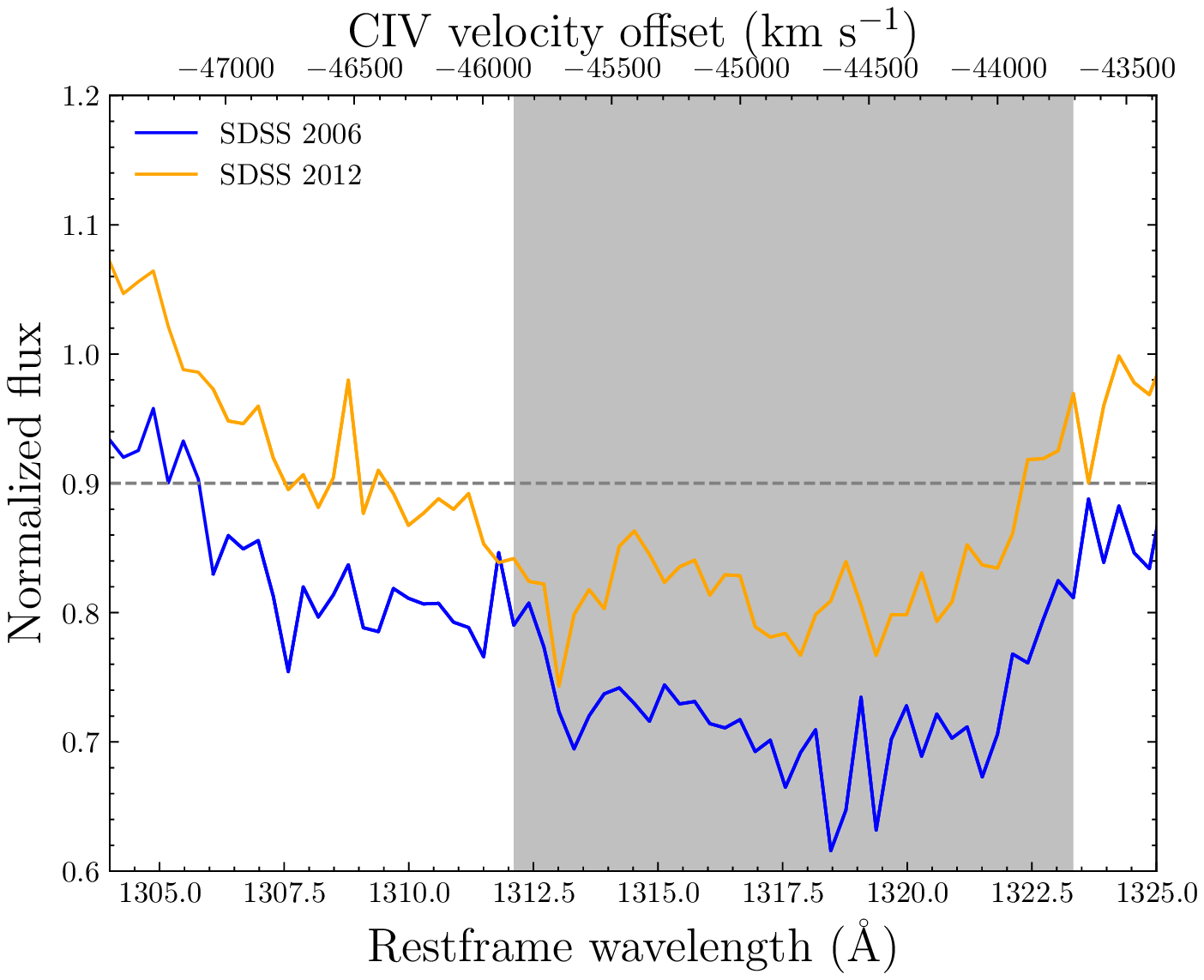}
         \end{minipage}%
 \begin{minipage}{.5\textwidth}
   \centering
   \includegraphics[width=7cm]{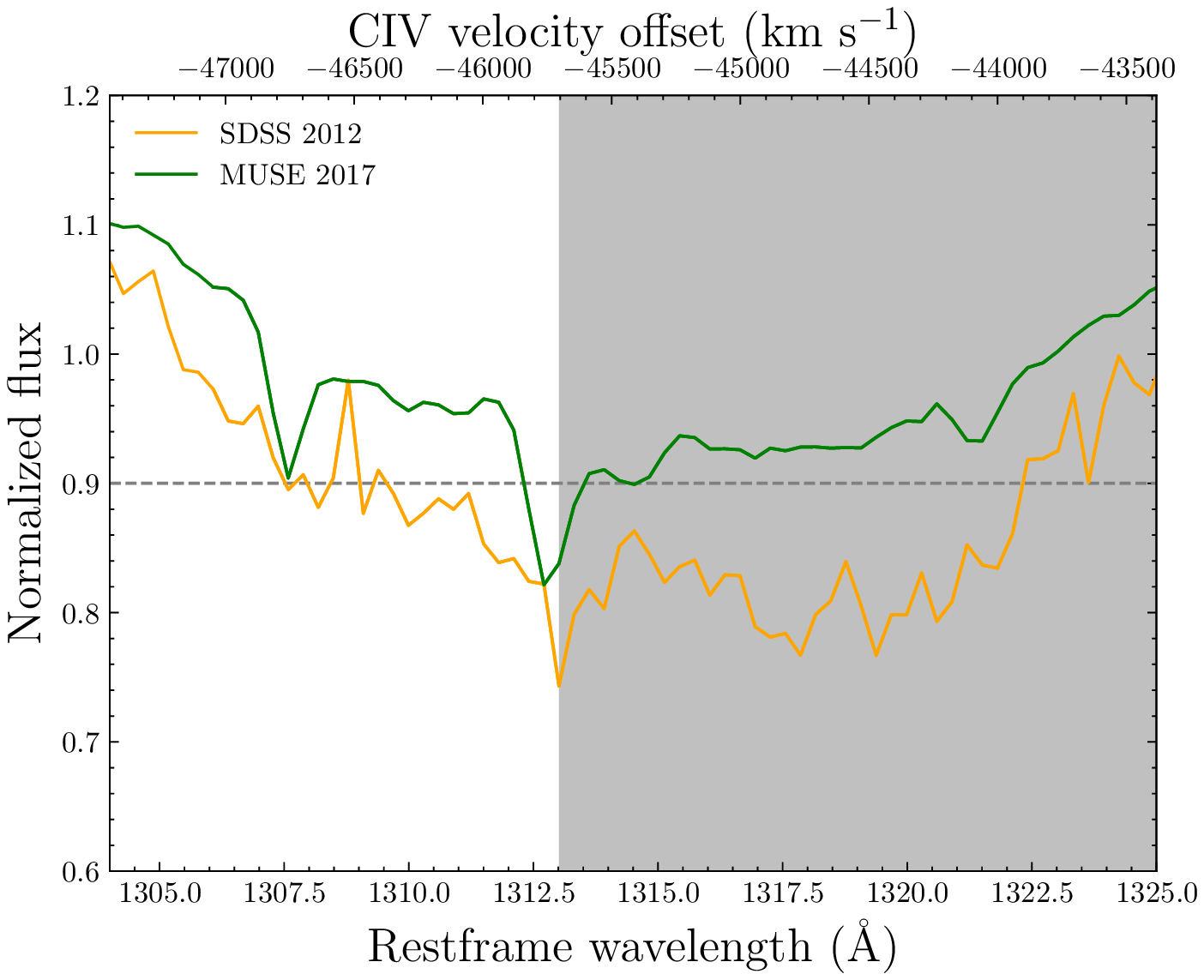}
         \end{minipage}\\
 \begin{minipage}{.5\textwidth}
   \centering
   \includegraphics[width=7cm]{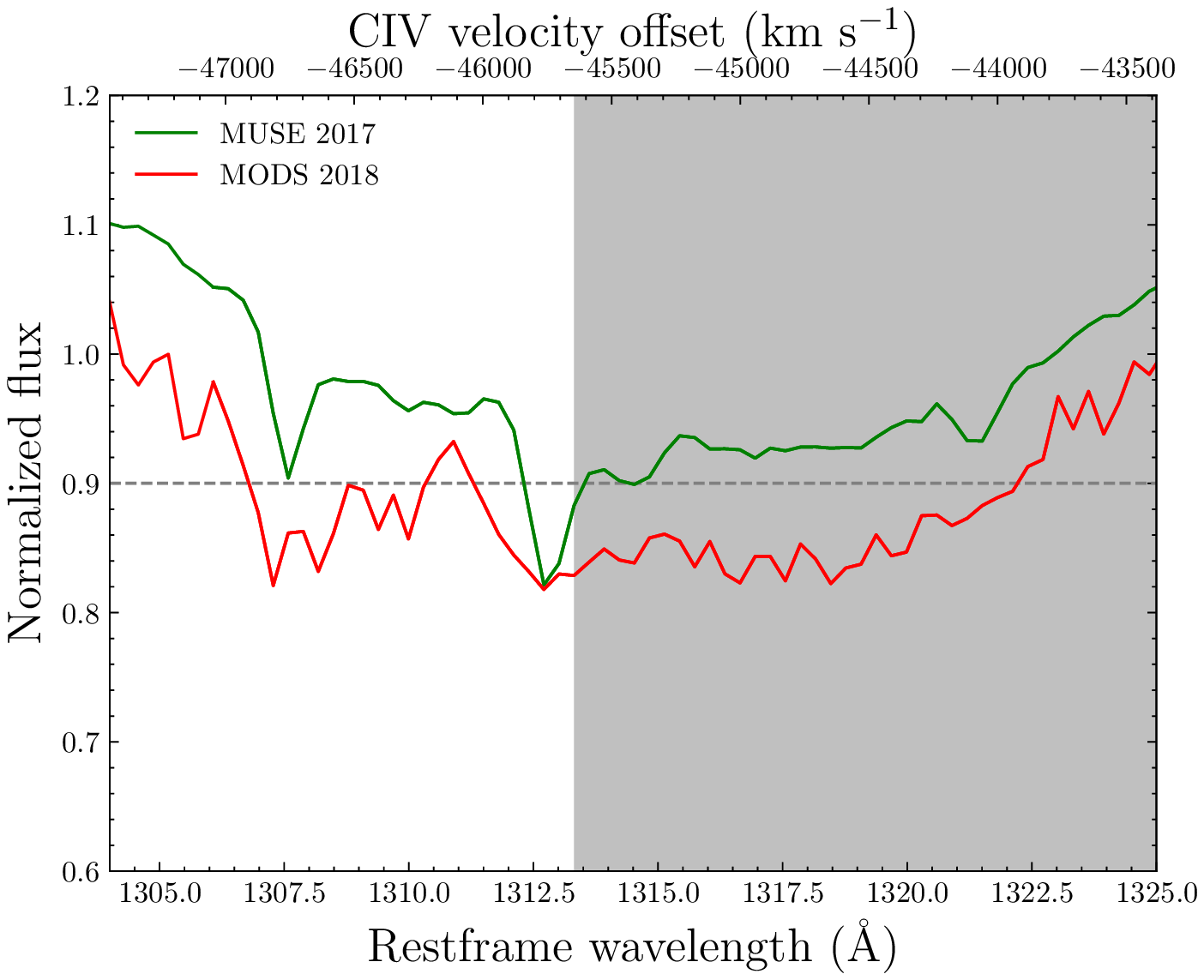}
         \end{minipage}%
 \begin{minipage}{.5\textwidth}
   \centering
   \includegraphics[width=7cm]{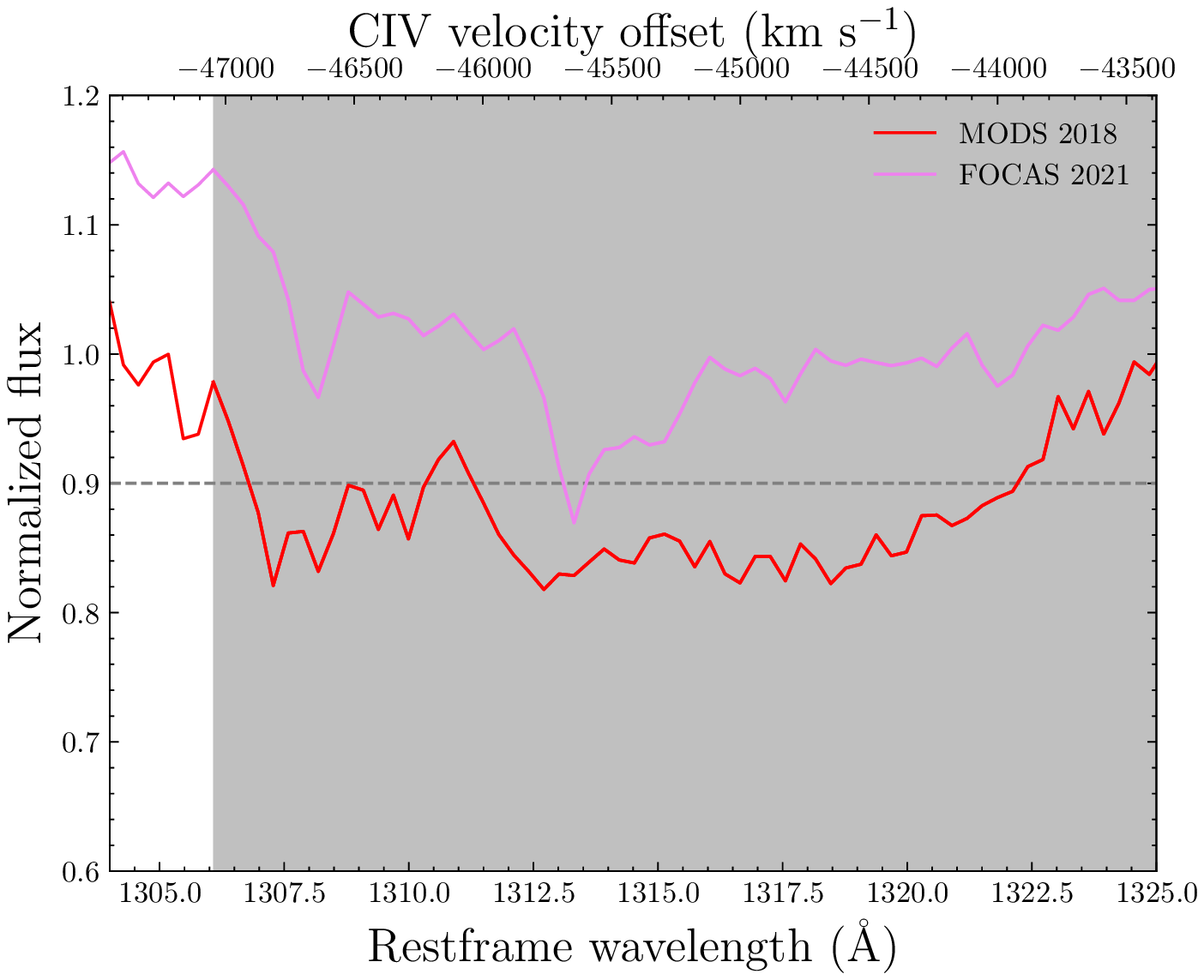}
        \end{minipage}
 \captionof{figure}{Comparison of spectra taken at two consecutive epochs over the velocity interval of the CIV B trough. Shaded region marks where the spectrum varies between the two epochs, which is used to define the region within the \textit{A$_s$} absorption strength is measured. Dashed line represents the 90\% level of the continuum-normalized flux.}\label{fig:D1_B}
         \end{minipage}%
         
         \newpage
         
 \noindent\begin{minipage}{\textwidth}
 \begin{minipage}{.5\textwidth}
   \centering
   \includegraphics[width=7cm]{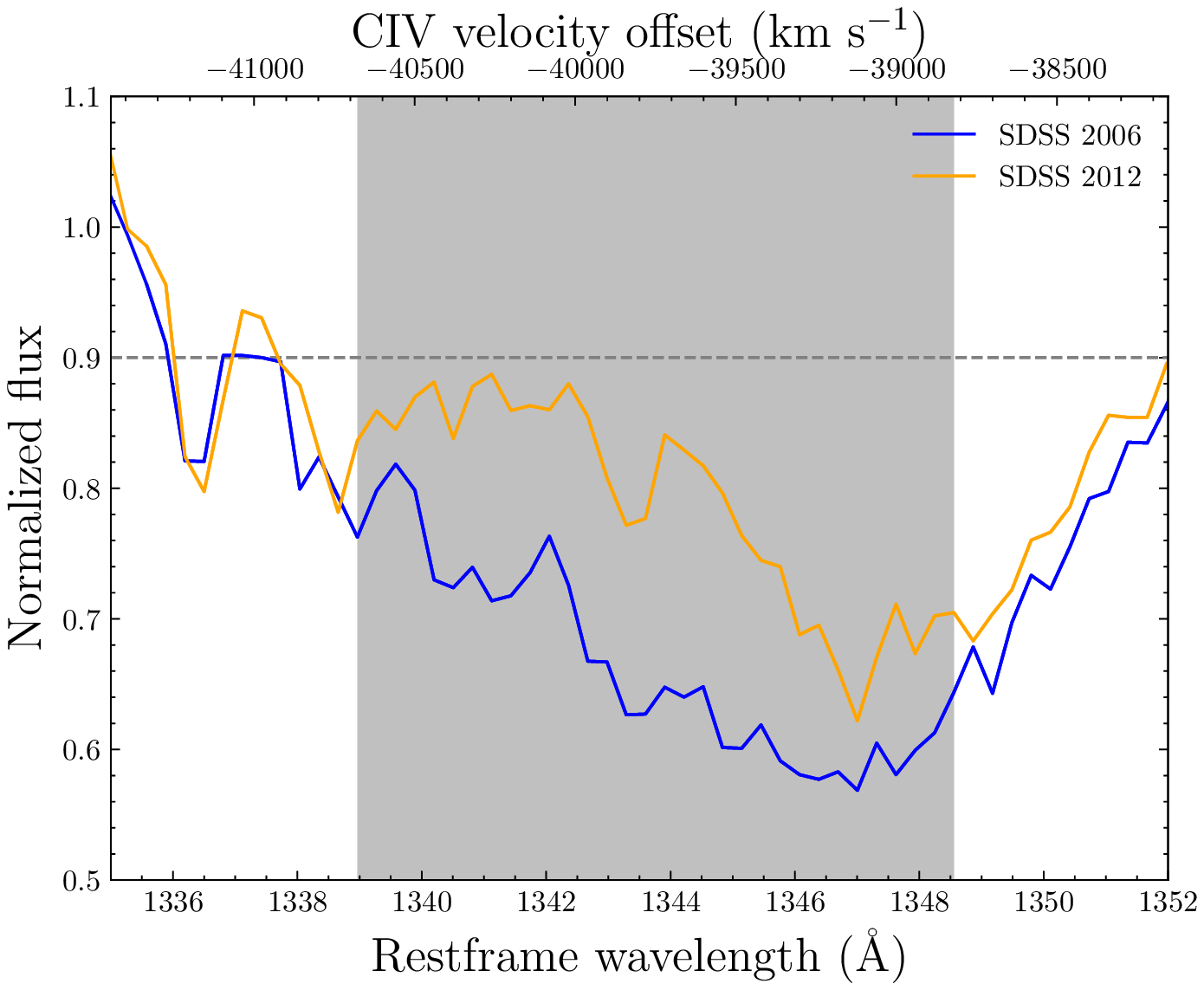}
         \end{minipage}%
 \begin{minipage}{.5\textwidth}
   \centering
   \includegraphics[width=7cm]{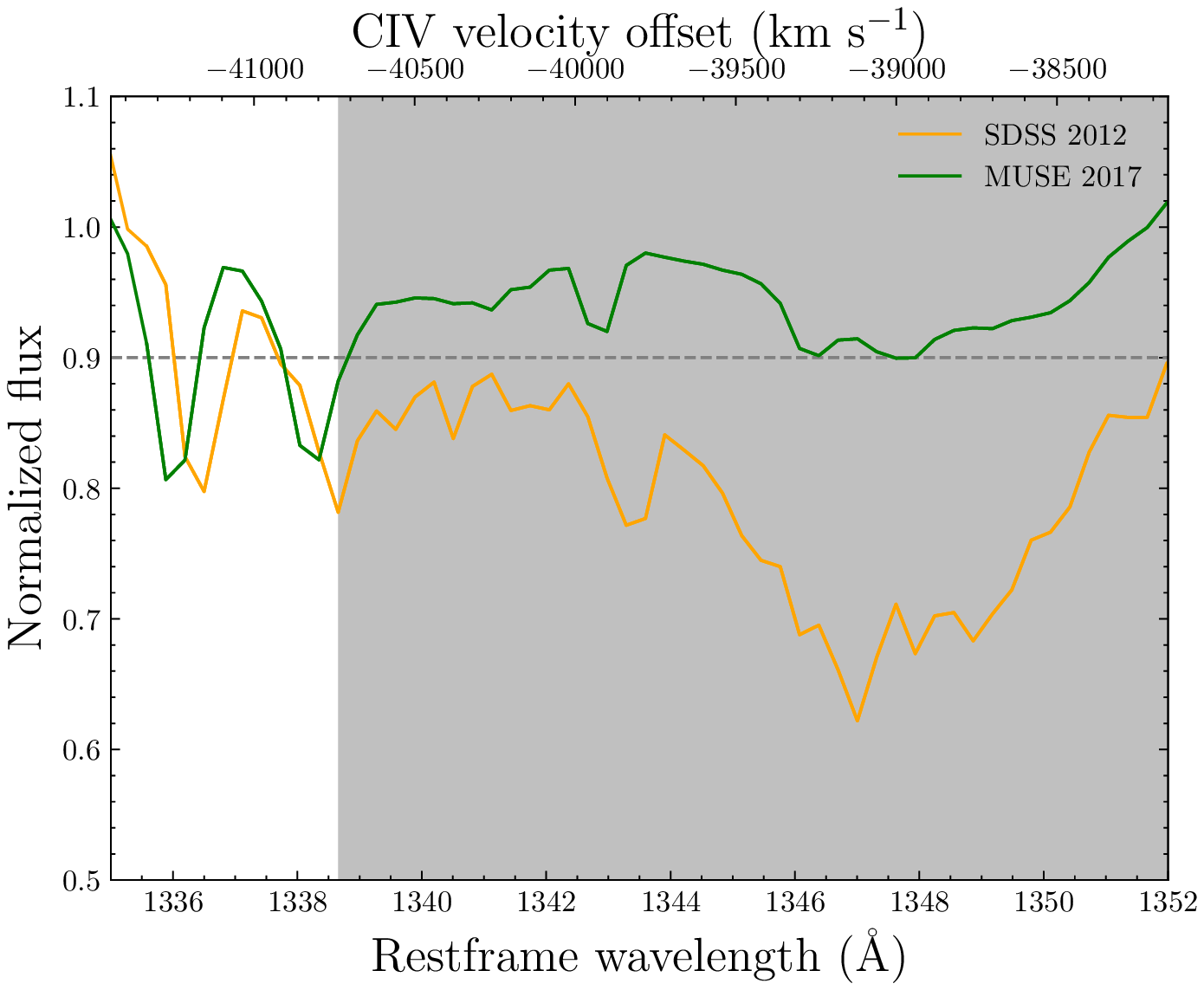}
         \end{minipage}\\
 \begin{minipage}{.5\textwidth}
   \centering
   \includegraphics[width=7cm]{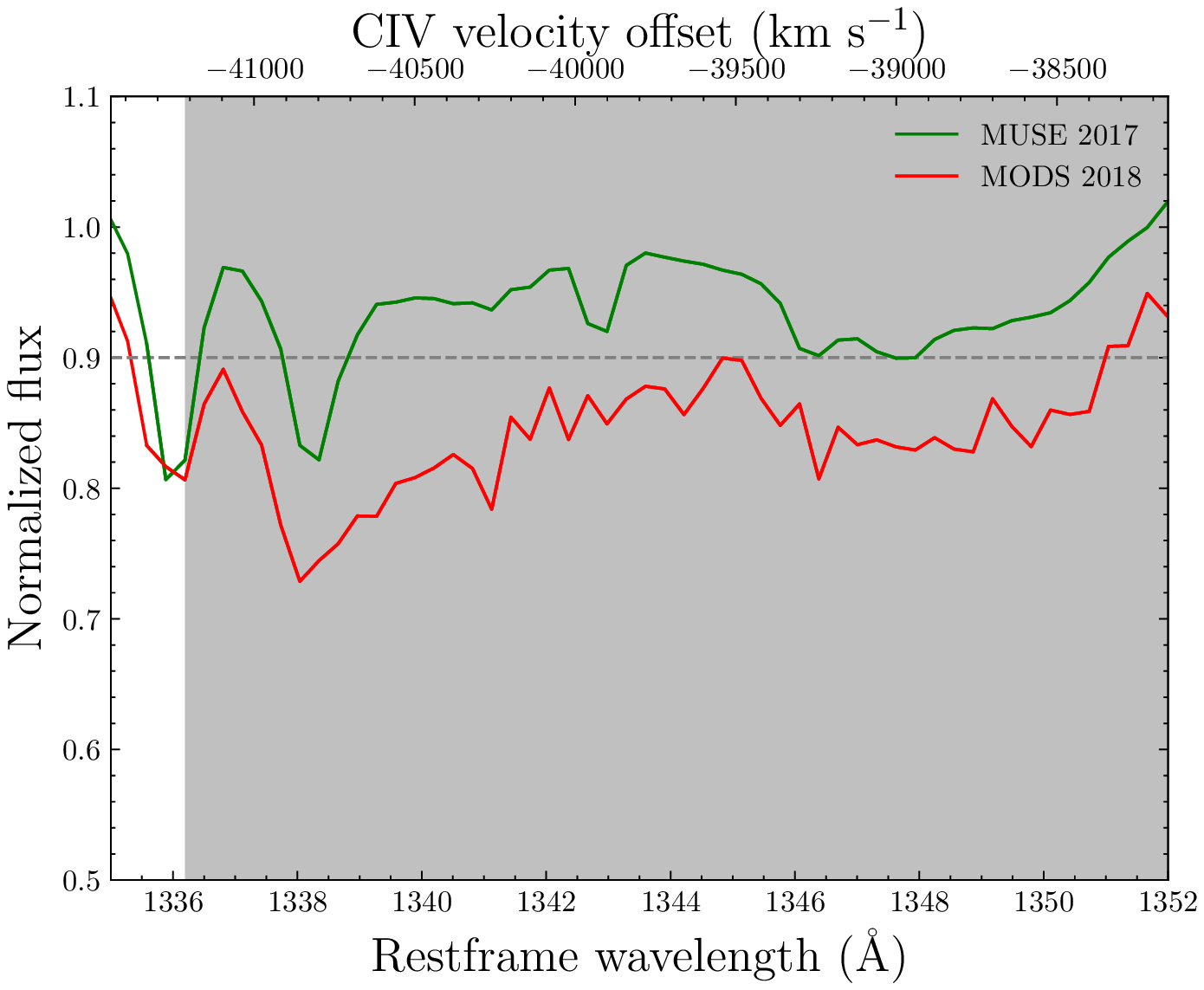}
         \end{minipage}%
 \begin{minipage}{.5\textwidth}
   \centering
   \includegraphics[width=7cm]{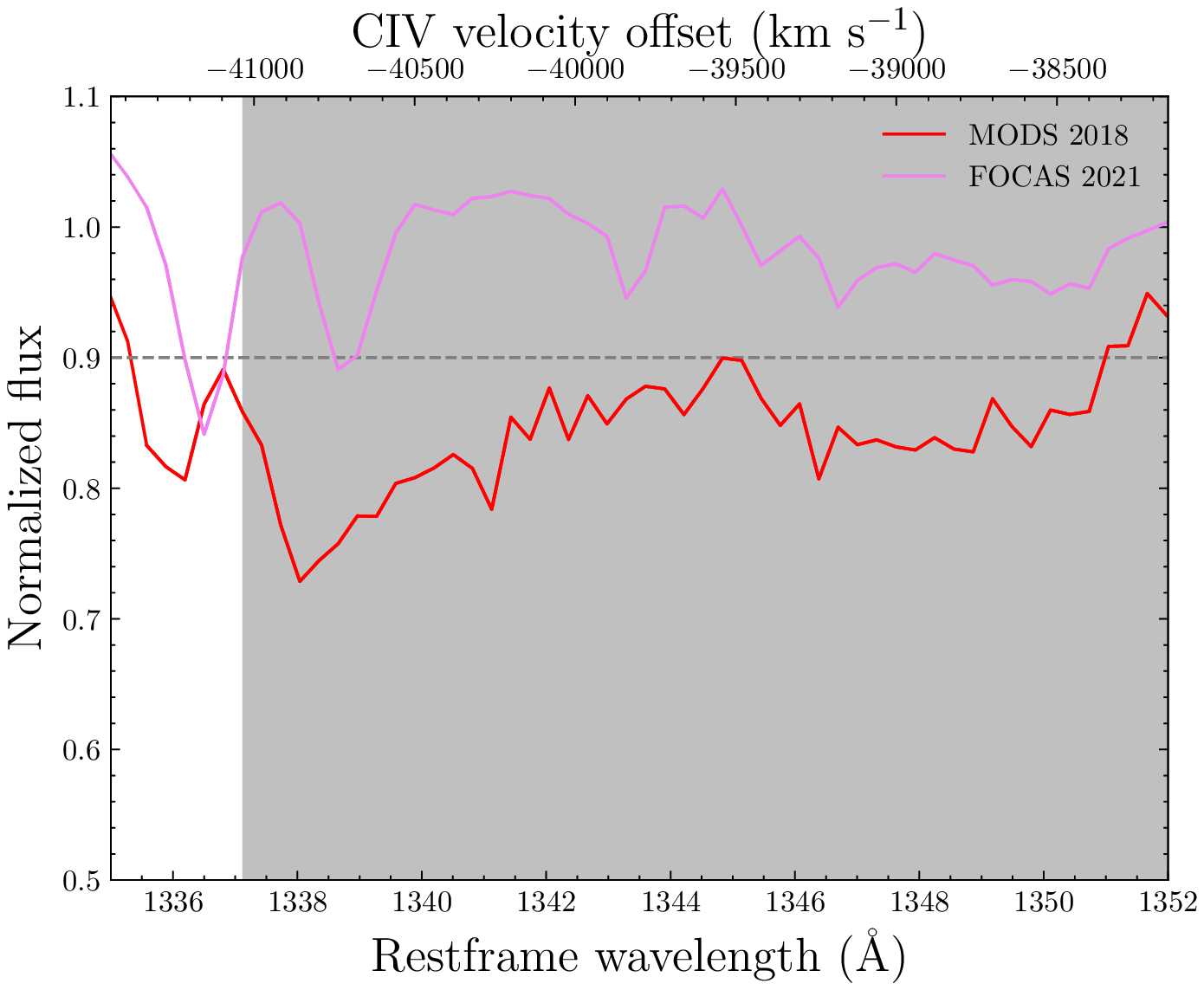}
         \end{minipage}%
 \captionof{figure}{Comparison of spectra taken at two consecutive epochs over the velocity interval of the CIV C trough. Shaded region marks where the spectrum varies between the two epochs, which is used to define the region within the \textit{A$_s$} absorption strength is measured. Dashed line represents the 90\% level of the continuum-normalized flux.}\label{fig:D1_C}
\end{minipage}%

\newpage

\end{document}